\documentclass[twocolumn]{aastex63}
\usepackage{booktabs}
\usepackage{natbib}
\received{\today}
\revised{--}
\accepted{--}

\shorttitle{2022crv}
\shortauthors{Gangopadhyay}
\begin{document}
	
	%\title{SN~2022crv: a type IIb SN with thin Hydrogen envelope}
	\title{Bridging between type IIb and Ib supernovae: SN IIb 2022crv with a very thin Hydrogen envelope}
	
	\correspondingauthor{Anjasha Gangopadhyay}
	\email{anjashagangopadhyay@gmail.com}
	
	\author[0000-0002-3884-5637]{Anjasha Gangopadhyay}
	\affiliation{Hiroshima Astrophysical Science Centre, Hiroshima University, 1-3-1 Kagamiyama, Higashi-Hiroshima, Hiroshima 739-8526, Japan}
	
	\author[0000-0003-2611-7269]{Keiichi Maeda}
	\affiliation{Department of Astronomy, Kyoto University, Kitashirakawa-Oiwake-cho, Sakyo-ku, Kyoto 606-8502, Japan}
	
	\author[0000-0003-2091-622X]{Avinash Singh}
	\affiliation{Hiroshima Astrophysical Science Centre, Hiroshima University, 1-3-1 Kagamiyama, Higashi-Hiroshima, Hiroshima 739-8526, Japan}
	
	\author[0000-0002-8070-5400]{Nayana A.J.}
	\affiliation{Indian Institute of Astrophysics, Koramangala 2nd Block, Bangalore 560034, India}
	
	\author{Tatsuya Nakaoka}
	\affiliation{Hiroshima Astrophysical Science Centre, Hiroshima University, 1-3-1 Kagamiyama, Higashi-Hiroshima, Hiroshima 739-8526, Japan}
	
	\author[0000-0001-6099-9539]{Koji S Kawabata}
	\affiliation{Hiroshima Astrophysical Science Centre, Hiroshima University, 1-3-1 Kagamiyama, Higashi-Hiroshima, Hiroshima 739-8526, Japan}
	\affiliation{Department of Physics, Graduate School of Advanced Science and Engineering, Hiroshima University, 1-3-1 Kagamiyama, Higashi-Hiroshima, Hiroshima 739-8526, Japan}
	
	\author[0000-0002-8482-8993]{Kenta Taguchi}
	\affiliation{Department of Astronomy, Kyoto University, Kitashirakawa-Oiwake-cho, Sakyo-ku, Kyoto 606-8502, Japan}
	
	\author[0000-0001-6706-2749]{Mridweeka Singh}
	\affiliation{Indian Institute of Astrophysics, Koramangala 2nd Block, Bangalore 560034, India}
	
	\author[0000-0002-0844-6563]{Poonam Chandra}
	\affiliation{National Radio Astronomy Observatory, 520 Edgemont Road, Charlottesville, VA, 22903, USA}

	\author[0000-0003-4501-8100]{Stuart D Ryder}
	\affiliation{School of Mathematical and Physical Sciences, Macquarie University, NSW 2109, Australia}
	\affiliation{Astrophysics and Space Technologies Research Centre, Macquarie University, Sydney, NSW 2109, Australia}
	
	\author[0000-0001-6191-7160]{Raya Dastidar}
	\affiliation{Instituto de Astrofísica, Universidad Andres Bello, Fernandez Concha 700, Las Condes, Santiago RM, Chile}
	\affiliation{Millennium Institute of Astrophysics, Nuncio Monsenor Sótero Sanz 100, Providencia, Santiago, 8320000 Chile}
	
	\author[0000-0001-9456-3709]{Masayuki Yamanaka}
	\affiliation{Amanogawa Galaxy Astronomy Research Center (AGARC), Graduate School of Science and Engineering, Kagoshima University, 1-21-35 Korimoto, Kagoshima, Kagoshima 890-0065, Japan}
	
	\author[0000-0002-4540-4928]{Miho Kawabata}
	\affiliation{Nishi-Harima Astronomical Observatory, Center for Astronomy,
		University of Hyogo, 407-2 Nishigaichi, Sayo-cho, Sayo, Hyogo, 679-5313, Japan}
	
	\author[0000-0001-5609-7372]{Rami Z. E. Alsaberi}
	\affiliation{Western Sydney University, Locked Bag 1797, Penrith, NSW, 2751, Australia}
	
	\author{Naveen Dukiya}
	\affiliation{Aryabhatta Research Institute of observational sciencES, Manora Peak, Nainital 263 002 India}
	\affiliation{Department of Applied Physics, Mahatma Jyotiba Phule Rohilkhand University, Bareilly, 243006, India}
	
	\author[0000-0002-0525-0872]{Rishabh Singh Teja}
	\affiliation{Indian Institute of Astrophysics, Koramangala 2nd Block, Bangalore 560034, India}
	\affiliation{Pondicherry University, Chinna Kalapet, Kalapet, Puducherry 605014, India}
	
	\author{Bhavya Ailawadhi}
	\affiliation{Aryabhatta Research Institute of observational sciencES, Manora Peak, Nainital 263 002 India}
	\affiliation{Department of Physics, Deen Dayal Upadhyaya Gorakhpur University, Gorakhpur, 273009, India}
	
	\author[0000-0002-7708-3831]{Anirban Dutta}
	\affiliation{Indian Institute of Astrophysics, Koramangala 2nd Block, Bangalore 560034, India}
	\affiliation{Pondicherry University, Chinna Kalapet, Kalapet, Puducherry 605014, India}
	
	\author[0000-0002-6688-0800]{D.K. Sahu}
	\affiliation{Indian Institute of Astrophysics, Koramangala 2nd Block, Bangalore 560034, India}
	
	\author[0000-0003-1169-1954]{Takashi J Moriya}
	\affiliation{National Astronomical Observatory of Japan, National Institutes of Natural Sciences, 2-21-1 Osawa, Mitaka, Tokyo 181-8588, Japan}
	\affiliation{School of Physics and Astronomy, Faculty of Science, Monash University, Clayton, Victoria 3800, Australia}
	
	\author[0000-0003-1637-267X]{Kuntal Misra}
	\affiliation{Aryabhatta Research Institute of observational sciencES, Manora Peak, Nainital 263 002 India}
	
	\author[0000-0001-8253-6850]{Masaomi Tanaka}
	\affiliation{Astronomical Institute, Tohoku University, Aoba, Sendai 980-8578, Japan}
	
	\author[0000-0002-9117-7244]{Roger Chevalier}
	\affiliation{University Of Virginia, Astronomy Building, 530 McCormick Road, Charlottesville VA 22904}
	
	\author[0000-0001-8537-3153]{Nozomu Tominaga}
	\affiliation{National Astronomical Observatory of Japan, National Institutes of Natural Sciences, 2-21-1 Osawa, Mitaka, Tokyo 181-8588, Japan}
	\affiliation{Astronomical Science Program, Graduate Institute for Advanced Studies, SOKENDAI, 2-21-1 Osawa, Mitaka, Tokyo 181-8588, Japan}
	\affiliation{Department of Physics, Faculty of Science and Engineering, Konan University, 8-9-1 Okamoto, Kobe, Hyogo 658-8501, Japan }
	
	\author[0000-0002-6765-8988]{Kohki Uno}
	\affiliation{Department of Astronomy, Kyoto University, Kitashirakawa-Oiwake-cho, Sakyo-ku, Kyoto 606-8502, Japan}
	
	\author{Ryo Imazawa}
	\affiliation{Department of Physics, Graduate School of Advanced Science and Engineering, Hiroshima University, Kagamiyama, 1-3-1 Higashi-Hiroshima, Hiroshima 739-8526, Japan}
	\affiliation{Hiroshima Astrophysical Science Centre, Hiroshima University, 1-3-1 Kagamiyama, Higashi-Hiroshima, Hiroshima 739-8526, Japan}
	
	\author{Taisei Hamada}
	\affiliation{Department of Physics, Graduate School of Advanced Science and Engineering, Hiroshima University, Kagamiyama, 1-3-1 Higashi-Hiroshima, Hiroshima 739-8526, Japan}
	\affiliation{Hiroshima Astrophysical Science Centre, Hiroshima University, 1-3-1 Kagamiyama, Higashi-Hiroshima, Hiroshima 739-8526, Japan}
	
	\author{Tomoya Hori}
	\affiliation{Department of Physics, Graduate School of Advanced Science and Engineering, Hiroshima University, Kagamiyama, 1-3-1 Higashi-Hiroshima, Hiroshima 739-8526, Japan}
	\affiliation{Hiroshima Astrophysical Science Centre, Hiroshima University, 1-3-1 Kagamiyama, Higashi-Hiroshima, Hiroshima 739-8526, Japan}
	
	\author{Keisuke Isogai}
	\affiliation{Okayama Observatory, Kyoto University, 3037-5 Honjo, Kamogatacho, Asakuchi, Okayama 719-0232, Japan}
	\affiliation{Department of Multi-Disciplinary Sciences, Graduate School of Arts and Sciences, The University of Tokyo, 3-8-1 Komaba, Meguro, Tokyo 153-8902, Japan}

	%% automatically handles all commas and "and"s between authors' names.
	
	%% can be used before or after the list of corresponding authors. The
	
	\begin{abstract}
		
		We present optical, near-infrared, and radio observations of supernova (SN) SN~IIb 2022crv. We show that it retained a very thin H envelope and transitioned from a SN~IIb to a SN~Ib; prominent H$\alpha$ seen in the pre-maximum phase diminishes toward the post-maximum phase, while He {\sc i} lines show increasing strength. \texttt{SYNAPPS} modeling of the early spectra of SN~2022crv suggests that the absorption feature at 6200\,\AA\ is explained by a substantial contribution of H$\alpha$ together with Si {\sc ii}, as is also supported by the velocity evolution of H$\alpha$. The light-curve evolution is consistent with the canonical stripped-envelope supernova subclass but among the slowest. The light curve lacks the initial cooling phase and shows a bright main peak (peak M$_{V}$=$-$17.82$\pm$0.17 mag), mostly driven by radioactive decay of $\rm^{56}$Ni. The light-curve analysis suggests a thin outer H envelope ($M_{\rm env} \sim$0.05 M$_{\odot}$) and a compact progenitor (R$_{\rm env}$ $\sim$3 R$_{\odot}$). An interaction-powered synchrotron self-absorption (SSA) model can reproduce the radio light curves with a mean shock velocity of 0.1c. The mass-loss rate is estimated to be in the range of (1.9$-$2.8) $\times$ 10$^{-5}$ M$_{\odot}$ yr$^{-1}$ for an assumed wind velocity of 1000 km s$^{-1}$, which is on the high end in comparison with other compact SNe~IIb/Ib. SN~2022crv fills a previously unoccupied parameter space of a very compact progenitor, representing a beautiful continuity between the compact and extended progenitor scenario of SNe~IIb/Ib.
		
	\end{abstract}
	
	
	\section{Introduction}
	\label{1}
	
	Core-collapse supernovae (CCSNe) show a great diversity in their observational properties, which reflects the diverse nature of their massive progenitor stars \citep[M\,$\gtrsim$\,8 M$_{\odot}$;][]{2003ApJ...591..288H,2009ARA&A..47...63S}, especially in the advanced phases of their evolution which is still poorly understood. This is highlighted by the so-called stripped-envelope supernovae (SE-SNe); they constitute a distinct class of CCSNe that strips off some or all of their outer H envelope. The mechanism for the envelope stripping is still controversial, with strong stellar winds \citep{2008A&ARv..16..209P} and/or interaction with a binary companion \citep{1992PASP..104..717P,2019NatAs...3..434F} being suggested. Therefore, SE-SNe have been intensively studied to understand the evolution of massive stars in their final phases. One important issue here is where the boundary between SNe~IIb and SNe~Ib/c (defined by the presence or absence of the H lines in their spectra, respectively) might lie concerning the nature of their progenitors.  
	
	The progenitors of SNe~IIb and SNe~Ib are believed to differ by the amount of the envelope stripping; if the majority of the H envelope is removed before the SN explosion, an H-free SN~Ib/c will be the outcome. In some cases, the progenitors of SE-SNe are identified in high-resolution images obtained by the Hubble Space Telescope and/or other facilities \citep{2015ApJ...811..147F,2018AAS...23232009M}. The methodology has been specially established for the class of SNe~IIb, which shows a diverse nature in their progenitors: for SNe~IIb 1993J and 2001ig, a supergiant in a binary interacting system is thought to be the progenitor \citep{2009Sci...324..486M,2018ApJ...856...83R}; the progenitor of SN~IIb 2008ax is likely a highly-stripped star (a low-mass analog of a Wolf-Rayet star) in a binary system with M$\rm_{ZAMS}\,\sim\,$ 10 -- 14 M$_{\odot}$ \citep{2008MNRAS.389..955P, 2008MNRAS.391L...5C}; while yellow supergiant progenitors with M$\rm_{ZAMS}\,\sim\,$ 10 -- 17 M$_{\odot}$ have been reported for SNe~IIb 2011dh, 2013df and 2016gkg \citep{2011ApJ...739L..37M, 2014ApJ...793L..22F, 2012ApJ...757...31B, maeda2015,2022ApJ...936..111K, 2022ApJ...934..186N}. On the other hand, the direct detection of the SN~Ib/Ic progenitors is far more challenging, so far resulting in only the three cases of iPTF13bvn, SN~2017ein, and SN~2019yvr whose progenitor masses were found between 10 -- 20 M$_{\odot}$ \citep{2013A&A...558L...1G}, 47 -- 80 M$_{\odot}$ \citep{2016MNRAS.461L.117E, 2018MNRAS.480.2072K}, and $\sim$ 10 M$_{\odot}$ \citep{2022MNRAS.510.3701S}, respectively. 
	
	Indeed, the absence of H in spectra of SNe~Ib does not necessarily mean that their progenitors are totally H-free. A question then is how much H envelope is required for SE-SNe to be classified as a SN~IIb? The presence of only a small amount of the H envelope prior to core collapse can lead to strong H features in the spectra during the photospheric phase, with the H$\alpha$ and H$\beta$ lines being the most prominent features, along with several strong He lines. The H features in SNe~IIb fade over time until the spectra become similar to those of SNe~Ib \citep{1988AJ.....96.1941F,2000hst..prop.8754F}. From their synthetic spectra, \citet{2012MNRAS.422...70H} concluded that even 0.025 $-$ 0.033 M$_{\odot}$ of the H mass can produce a strong H$\alpha$ absorption feature, suggesting that there is a blending between SNe~IIb and SNe~Ib.\citet{2017ApJ...840...10Y} and \citet{2019ApJ...885..130S} also showed that if the H mass remains between 0.001 M$_{\odot}$ and 0.5 M$_{\odot}$, the features of SNe~IIb will arise. \cite{2022MNRAS.511..691G} recently showed that the minimum mass threshold of H for a SNe~IIb is 0.033 M$_{\odot}$ similar to \citet{2012MNRAS.422...70H}, but inconsistent with \citet{2019ApJ...885..130S}.  \cite{2017MNRAS.469.2672P} proposed two additional SE-SNe subcategories: the SNe~IIb(I), showing moderately H-rich spectra in which the H$\alpha$ P-Cygni profile is dominated by the absorption component relative to the emission profile; and the SNe~Ib(II), showing weak residual H$\alpha$, but no clear appearance of other Balmer lines. The findings by \cite{2017MNRAS.469.2672P} along with \cite{2012MNRAS.422...70H} indicate that SNe~IIb and SNe~Ib are linked more physically than thought before. 
	
	The nature of the SE-SN progenitors can also be obtained from their light curve properties; the `compact' SNe~IIb have a similar bolometric light curve shape to SNe~Ib, while `extended' SNe~IIb show a double-peaked light curve owing to a combination of the shock-cooling emission and the radioactive decay of $^{56}$Ni to $^{56}$Co (and then to $^{56}$Fe) \citep{2004Natur.427..129M,2014MNRAS.445.1647M}. Except for the extended SNe~IIb, the first peak due to the shock cooling is not observed in most SE-SNe because the progenitor compactness causes the shock-cooling emission to decay too quickly. The typical rise times of SE-SNe lie in the range of 10--20 d \citep{2019MNRAS.485.1559P} with the average peak absolute magnitudes of M$_B$ $\sim$ $-$16.99$\pm$0.45 and $-$17.66$\pm$0.40 mag \citep{2006AJ....131.2233R,2014AJ....147..118R} for SNe IIb and Ib, respectively. 
	
	Another powerful method to constrain the nature of the progenitors is radio observation. Radio emission from SE-SNe results from an interaction between the SN shock wave and nearby circumstellar medium (CSM) \citep{1982ApJ...259..302C,1998ApJ...499..810C}. Radio observations help probe the density structure of the CSM, and thus the mass-loss history of the progenitor \citep{1986ApJ...301..790W,1982ApJ...259..302C,1998ApJ...499..810C,2021ApJ...918...34M}. Indeed, the `extended' and `compact' SNe IIb classification is linked to the radio property \citep{chevalier2010}. The extended and compact progenitors naturally arise by the difference in the mass of the H-rich envelope, which also explains the differences in the radio signal through differences in the mass-loss history (e.g., \citealp{maeda2015,2017ApJ...840...90O}), although this idea has been tested only for a small sample. 
	
	In this paper, we study the optical (spectroscopic and photometric), infrared, and radio evolution of SN~IIb 2022crv. SN~2022crv was discovered by the Distance Less Than 40 Mpc (DLT40) survey on 2022-02-19 UT 04:30:59.616 (JD~2459629.69) \citep{2022TNSTR.448....1D} using the 0.4m Prompt5 telescope at an unfiltered mag of $\sim$\,18.05 mag. The SN is located at $\mathrm{R.A.} = 09^\mathrm{h}54^\mathrm{m}25.890^\mathrm{s}$, $\mathrm{Dec} = -25\degr42^{\prime}11\farcs07$ (equinox J2000.0) \citep{2022TNSCR.454....1A} which is 36\farcs7 west and 1\farcs3 north from the core of the host galaxy NGC~3054.
	
	The estimated explosion epoch and basic parameters of the SN are noted in subsections~\ref{expepoch} and \ref{dist-extinction}. The optical and near-infrared data reduction and analysis procedures are described in subsection~\ref{opt}. A detailed description of the spectral evolution, velocity evolution, and \texttt{SYNAPPS} \citep{2013ascl.soft08007T} spectral modeling is given in Section~\ref{spec}. The photometric evolution is elaborated in Section~\ref{phot}, where we discuss the light curve, color evolution, and absolute magnitudes of SN~2022crv compared to other SE-SNe. The bolometric light curve modeling and parameters are described in Section~\ref{bol}. The radio data reduction and analysis are described in Section~\ref{radio}. We discuss the progenitor properties of SN~2022crv in Section~\ref{progenitor}. Finally, we summarise the results of the study in Section~\ref{sum}.
	
	\section{Observations and Data reduction}
	\label{observations}
	\subsection {Estimation of the explosion epoch}
	\label{expepoch}
	
	The last non-detection of SN~2022crv before discovery was reported to be 2022-02-17 UT 04:42:50 (JD~2459627.70) at a magnitude upper limit of 17.83 with the Prompt5 0.4m telescope by \citet{2022TNSTR.448....1D}. In addition, Asteroid Terrestrial-impact Last Alert System  (ATLAS) data \citep{2018tonry, 2020smith} reveal a deeper non-detection of the source on 2022-02-16 (JD~2459627.46) at a magnitude limit of 18.73 $\pm$ 0.40, which is indeed more useful than the later non-detection in constraining the explosion epoch. 
	
	We estimated the explosion epoch of SN~2022crv by fitting the very early bolometric evolution using the Valenti and Nagy-Vinko models (see Section \ref{bol}). The estimated values of the explosion epoch from these two methods are JD~2459627.8 $\pm$0.5 and JD~2459628.0$\pm$0.5, respectively. An average of the above two estimates, combined with the ATLAS non-detection, constrains the explosion time t$_{0}$ to be JD\,=\,2459627.75$\pm$0.5. With this explosion epoch, the rise time of SN 2022crv bf to reach the bolometric maximum is 15.2\,d which is consistent with those found for typical SE-SNe as derived by  \cite{2019MNRAS.485.1559P} (7.8\,--\,20.7 d to reach bolometric maximum) and \cite{2016MNRAS.457..328L} ($\sim$\,18 d in the $R$-band). 
	
	\subsection{Distance \& extinction} 
	\label{dist-extinction}
	
	Adopting $H_0\,=\,73$~km~s$^{-1}$~Mpc$^{-1}$, $\Omega_{\rm matter}$\,=\,0.27 and $\Omega_{\rm vacuum}$\,=\,0.73 \citep{2007ApJS..170..377S}, we obtain a distance of 34.4 $\pm$ 2.4 Mpc, corrected for Virgo, Shapley, and Great Attractor, to the host galaxy NGC~3054 (with a redshift of $z=0.008091$) as computed on the NASA/IPAC Extragalactic Database (NED)\footnote{\url{https://ned.ipac.caltech.edu/}}. The Milky Way (MW) extinction along the line of sight to SN~2022crv is $A_V$\,=\,0.205~mag \citep{2011ApJ...737..103S}. 
	
	We see a conspicuous dip of \ion{Na}{1}~D at 5891.5~\AA\ in the spectra of SN~2022crv taken on 2022-03-12 (JD 2459651.2), 2022-03-29 (JD 2459668.0), and 2022-04-08 (JD 2459678.0) UT. For estimating the extinction within the host galaxy, we measure equivalent widths (EW) of the \ion{Na}{1}~D line in the combined spectra of these three dates to increase the signal-to-noise ratio. Using the formulation by \cite{2012MNRAS.426.1465P}, we estimate the host galaxy extinction as $A_V$\,=\,0.467$\pm$0.010~mag. We thus adopt $A_V$\,=\,0.672$\pm$0.010 mag as a combination of the extinction within the host galaxy and that within the MW along the line of sight. We use these values of distance and extinction throughout the paper.
	
	\subsection{Optical and Near-Infrared Observations}
	\label{opt}
	
	We observed SN~2022crv with the \textit{BgVriRI} filters from $\sim$ $-$8 to 86 d with respect to the $V$-band maximum \footnote{All the epochs in the paper are defined with respect to the V-band maximum which occurs at JD~2459644.19 and is computed in Section~\ref{phot}.}. The imaging observations were carried out using the 1.5m Kanata telescope (\citealt{2008SPIE.7014E..4LK}; KT) of Hiroshima University, 3.8m Seimei Telescope \citep{2020PASJ...72...48K} of Kyoto University at Okayama observatory,  1m Sampurnanand Telescope (\citealt{1999A&AS..135..391S}; ST), 1.3m Devasthal Fast Optical Telescope (\citealt{2012SPIE.8444E..1TS}; DFOT), Aryabhatta Research Institute of observational sciencES (ARIES), India and 2m Himalayan Chandra Telescope (\citealt{2010ASInC...1..193P}; HCT), Indian Institute of Astrophysics (IIA), India.
	
	Several bias, dark, and twilight flat frames were obtained during the observing runs along with science frames. For the initial pre-processing, several steps, such as bias-subtraction, flat-fielding correction, and cosmic ray removal, were applied to raw images of the SN. We implemented the standard tasks available in the data reduction software IRAF\footnote{IRAF stands for Image Reduction and Analysis Facility distributed by the National Optical Astronomy Observatory, operated by the Association of Universities for Research in Astronomy (AURA) under a cooperative agreement with the National Science Foundation.} for carrying out the pre-processing. Multiple frames were taken on some nights and co-added in respective bands after the geometric alignment of the images to increase the signal-to-noise ratio.
	
	To calibrate the secondary standards in the SN field, we observed a set of Landolt equatorial standards \citep{1990AJ....100..695L}: PG~1323, PG~0942, SA~32, and SA~104 on 2022-05-19 (JD 2459718.5) and 2022-05-20 (JD 2459719.5) UT using the 1.3m DFOT and 2m HCT. The observed Landolt field stars with the magnitudes of 10 $\leq$ V $\leq$ 13 were observed in a typical seeing of $1\farcs5$. The average site extinction values in the {\it BVRI} bands were taken from \cite{2008BASI...36..111S}. We calibrated 13 non-variable local standards in the SN~field using the transformation equations. These secondary standards were used to convert the instrumental magnitudes into apparent magnitudes. The calibrated {\it BVRI} magnitudes of the secondary standards averaged over two nights are listed in Table \ref{tab:photstandard}. The Point Spread Function (PSF) photometry for the data from ST, DFOT, and HCT was implemented through a reduction pipeline built in Python called \textsc{RedPipe}\footnote{\url{https://github.com/sPaMFouR/RedPipe}} \citep{2021redpipe}. SN~magnitudes were calibrated using the nightly zero points obtained from the secondary standards. 
	
	For the data taken by TriCCS (TriColor CMOS Camera and Spectrograph) attached to the Seimei Telescope, the {\it gri} band observations were calibrated using the APASS catalog\footnote{\href{APASS Catalog}{\url{https://www.aavso.org/aavso-photometric-all-sky-survey-data-release-1}}} with the same set of secondary standards. The {\it gri} band magnitudes from the images taken by the Seimei Telescope were measured using DAOPHOT\footnote{Dominion Astrophysical Observatory Photometry}. The {\it ri} band magnitudes were then converted to {\it RI}. We further added supplemental data from ATLAS, using their forced-photometry archive\footnote{\url{https://fallingstar-data.com}} developed by \citet{2021shingles}. In addition, we added {\it g}-band data of SN~2022crv from the All Sky Automated Search for SuperNovae (ASAS-SN, \citealp{2014ApJ...788...48S,2017PASP..129j4502K}). These additional data points were merged with the light curve from the Seimei Telescope. The final SN~magnitudes from all the instrumental setups are tabulated in Table \ref{tab:photopt}.
	
	The near-infrared (NIR) data of SN~2022crv were obtained with the HONIR instrument of KT. The sky-background subtraction was done using a template sky image obtained by dithering individual frames at different positions. We performed PSF photometry and calibrated the SN magnitudes using comparison stars in the 2MASS catalog \citep{1998AJ....116.2475P}. The final NIR magnitudes in the SN field are shown in Table~\ref{tab:photnir}.
	
	The spectroscopic observations were carried out using the Himalayan Faint Object Spectrograph and Camera (HFOSC) mounted on the HCT, KOOLS-IFU \citep{2019PASJ...71..102M} on the Seimei Telescope, and Aries Faint Object Spectrograph and Camera (ADFOSC) \citep{2016SPIE.9908E..4YK} mounted on the 3.6m Devasthal Optical Telescope (DOT), ARIES, India. Our spectral coverage spans from $-$10 d to +33\,d. We included the publicly available Gemini-N/GMOS spectrum taken on $-$15 d in our analysis \citep{2022TNSCR.454....1A}. For HCT, we used a 2$\arcsec$ wide slit and Grisms Gr7/Gr8 for taking optical spectra. The DOT spectrum was taken with the 676R grism and similar slit size.  The spectra taken with HFOSC and ADFOSC were reduced using the \textit{twodspec} package in IRAF, followed by wavelength and flux calibration. The slit loss corrections were done by scaling the spectra with respect to the SN photometry. The spectra with KOOLS-IFU were taken through optical fibers and the VPH-blue grism. The data reduction was performed using the Hydra package in IRAF \citep{1994ASPC...55..130B} and a reduction software developed for KOOLS-IFU data\footnote{\url{http://www.o.kwasan.kyoto-u.ac.jp/inst/p-kools}}. For each frame, we performed sky subtraction using a sky spectrum created by combining fibers to which the contributions from the object are negligible. Arc lamps of Hg, Ne, and Xe were used for wavelength calibration. Finally, the spectra were corrected for the heliocentric redshift of the host galaxy. The log of spectroscopic observations is reported in Table~\ref{tab:2022crv_spec_obs}. 
	
	\section{Spectroscopic Evolution}
	\label{spec}
	
	\begin{figure}
		\centering   
		\resizebox{\hsize}{!}{\includegraphics{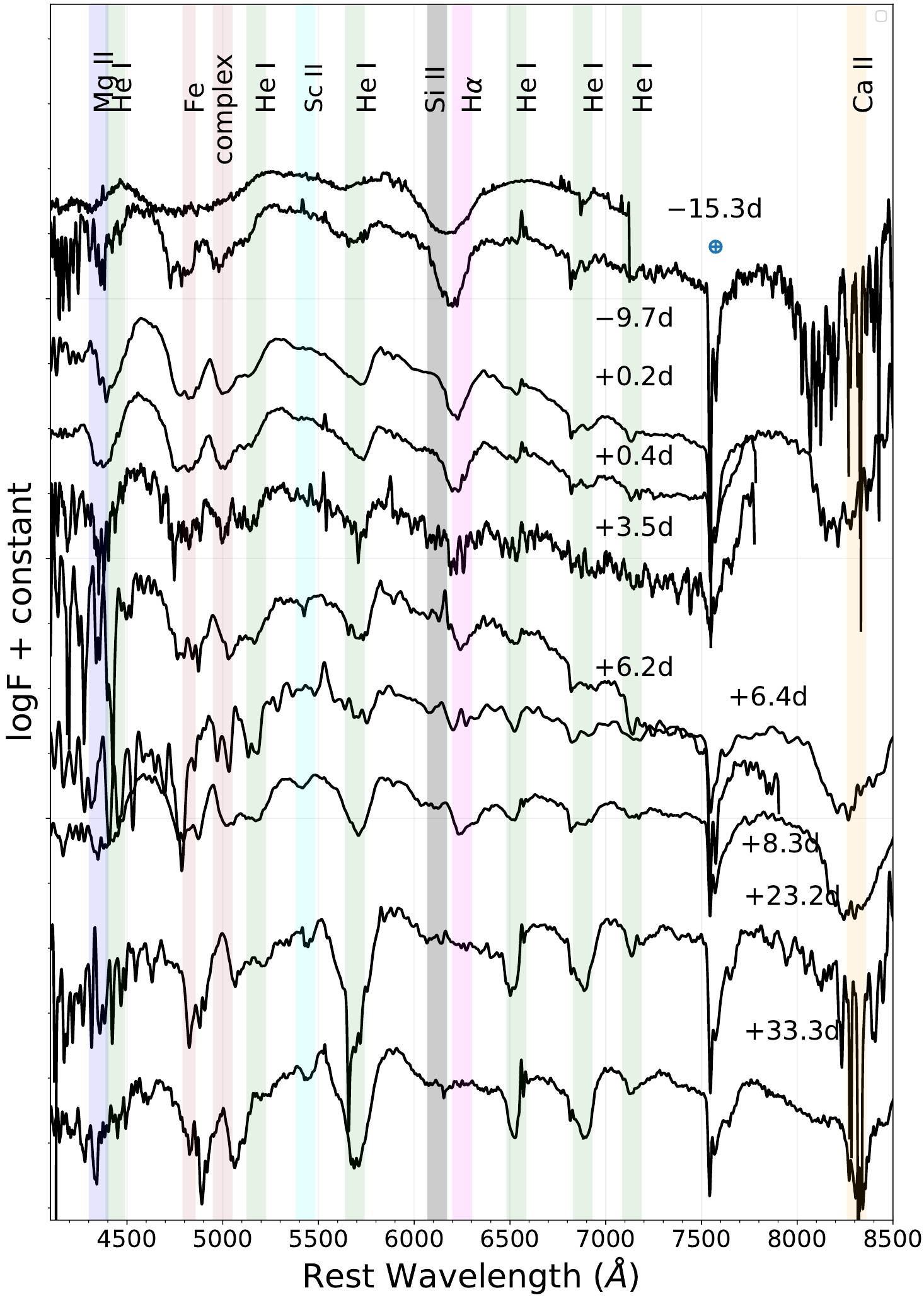}} 
		\caption{The complete spectral evolution of SN~2022crv beginning at $-$15.3 d, up to $+$33.3 d. The earliest spectrum shows a prominent H$\alpha$ dip which slowly diminishes around the maximum, and prominent lines of He {\sc i} start developing.}
		\label{fig:spec}
	\end{figure}
	
	Figure~\ref{fig:spec} shows the spectral evolution of SN~2022crv from $-$15.3 to 33.3 d. Our spectra mainly cover the early photospheric phase, which helps examine the properties of the outermost regions of the expanding ejecta.  
	The very early spectrum ($-$15.3 d) shows a broad absorption dip centered at 6170\,\AA, probably due to H$\alpha$ with some additional contribution from Si\,{\sc ii} 6355\,\AA, an absorption dip centered around $\sim$ 5700\,\AA\  due to He\,{\sc i} 5876\,\AA, and other features of Fe around 5000\,\AA. The absorption at 6170\,\AA\ corresponds to a high velocity of 18,000 km s$^{-1}$, if it is due to H$\alpha$. The second spectrum at $-$9.7 d shows features of the Fe {\sc ii} triplet (4924, 5018, 5169\,\AA), the He\,{\sc i} 5876\,\AA\ feature, and He\,{\sc i} 6678\,\AA\ superposed with the narrow H$\alpha$ from the host galaxy. Narrow emission lines due to the host galaxy are also seen for [N\,{\sc ii}] 6584\,\AA. We also see an absorption around 8498\,\AA\ due to the Ca\,{\sc ii} NIR triplet. The feature around 4500\,\AA\ spectra at +0.2 d and +0.4 d look very similar to Mg\,{\sc ii}. The feature appearing at 5500\,\AA\ is most likely a blend of Fe\,{\sc ii} at 5535\,\AA\ and Sc\,{\sc ii} at 5527\,\AA. The spectra during this phase show prominent absorption due to a combination of He\,{\sc i} 5876\,\AA\ and the Na\,{\sc i}\,D absorption. The spectra from +3.5 d to +33.3 d show that He {\sc i} 5876, He {\sc i} 6678, and He {\sc i} 7065\,\AA\ grow stronger over time. 
	
	The spectra also show a ``W"-shaped absorption feature centered around 5000\,\AA. A similar feature has been previously observed in SN~II 2005ap \citep{2007ApJ...666.1093Q}, SN~Ib 2008D \citep{2009ApJ...702..226M}, SN~Ib 2009jf \citep{2011MNRAS.413.2583S}, SN~IIb 2001ig \citep{2009PASP..121..689S} and SN~Ib 2015ap \citep{2020MNRAS.497.3770G}. For most of them, this spectral feature is typically seen in the pre-maximum time between $-$14 d to $-$10 d. \cite{2008Sci...321.1185M} explained the origin of this feature as Fe\,{\sc ii} complexes; this is confirmed for SN 2022crv by spectral modeling shown in later sections, but with an additional contribution by Mg {\sc ii}.

	\subsection{Spectral Comparison}
	\label{spec-comp}
	
	\begin{figure}
		\centering   
		\resizebox{\hsize}{!}{\includegraphics{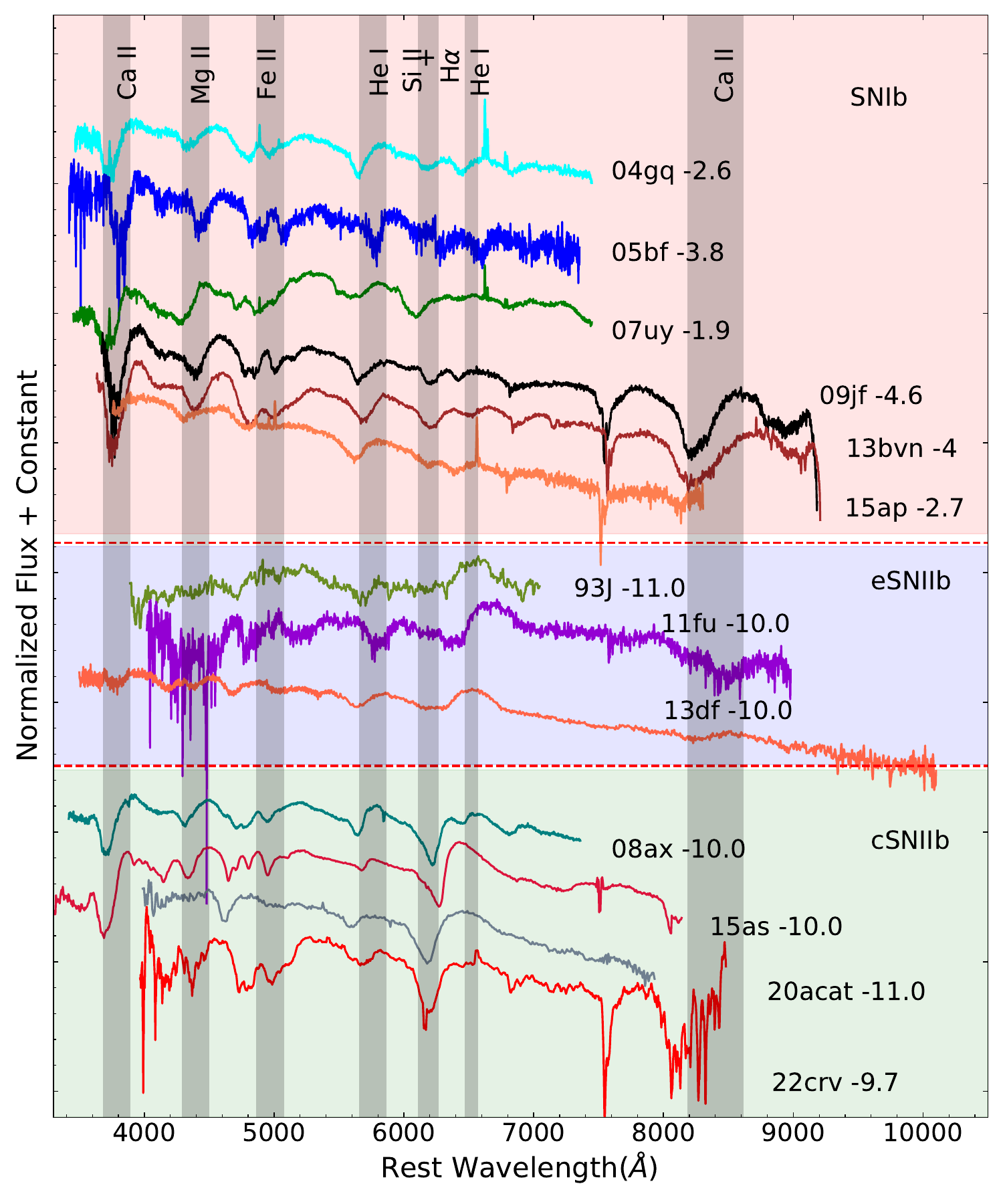}} 
		\caption{The pre-maximum spectrum of SN~2022crv as compared with other members of the SNe~IIb and Ib classes. The pink shaded region (top) shows the SN~Ib comparison sample, the blue shaded region (middle) shows the eSN~IIb, and the green shaded region (bottom) shows the cSNe~IIb comparison sample. The labels include the SN name plus the date since maximum light. The data for the comparison sample are taken from WiseRep and the papers given in Table~\ref{tab:compsample}.}
		\label{fig:precomp}
	\end{figure}
	
	\begin{figure}
		\centering   
		\resizebox{\hsize}{!}{\includegraphics{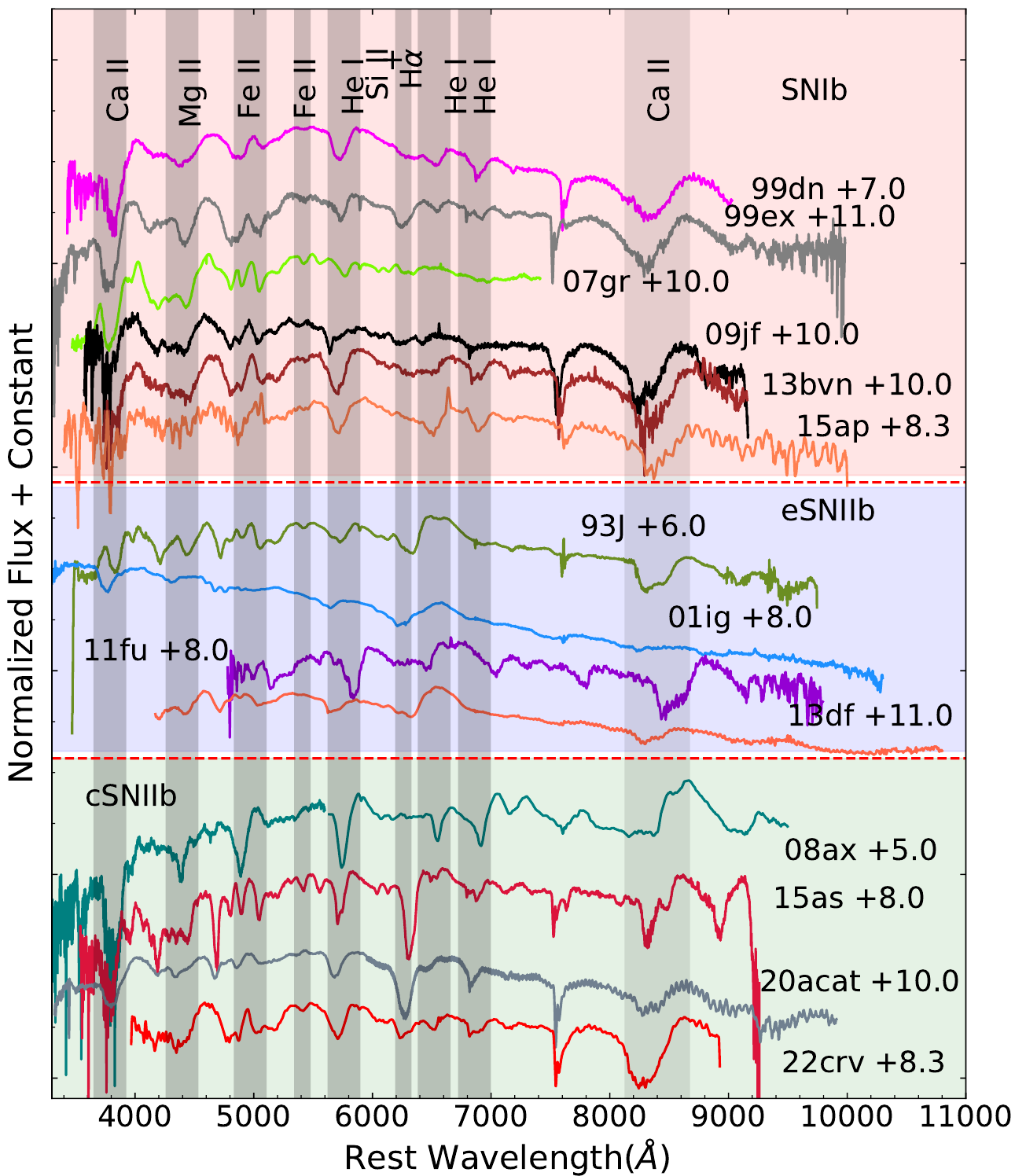}} 
		\caption{The post-maximum spectrum of SN~2022crv as compared with a group of SE-SNe. The layout and the comparison samples follow the descriptions in Figure~\ref{fig:precomp}.}
		\label{fig:10daycomp}
	\end{figure}
	
	\begin{figure}
		\centering   
		\resizebox{\hsize}{!}{\includegraphics{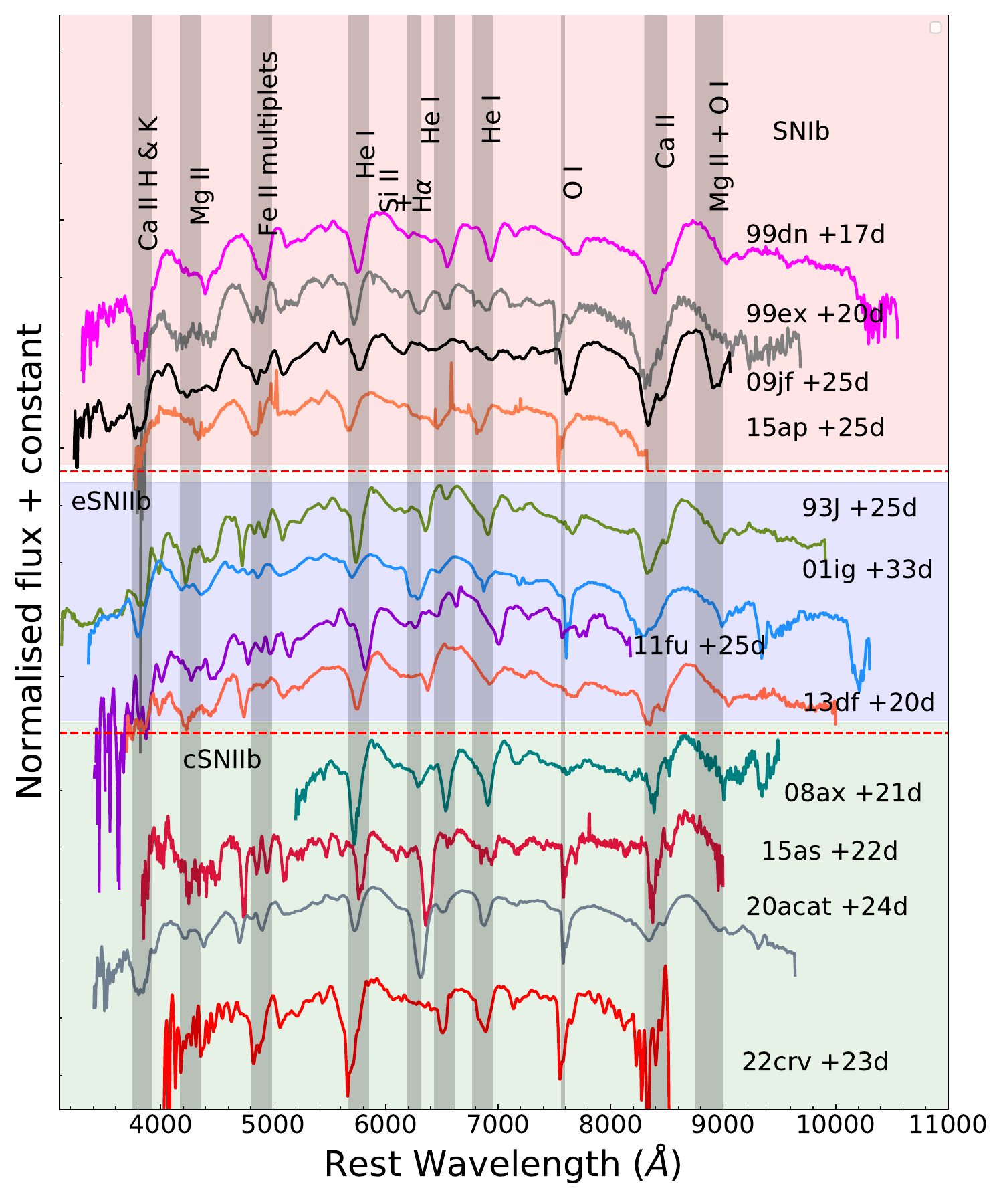}} 
		\caption{The late-time spectrum of SN~2022crv as compared with a group of SE-SNe. The layout and the comparison samples follow the descriptions in Figure~\ref{fig:precomp}.}
		\label{fig:25daycomp}
	\end{figure}
	
	\begin{figure}
		\centering   
		\resizebox{\hsize}{!}{\includegraphics{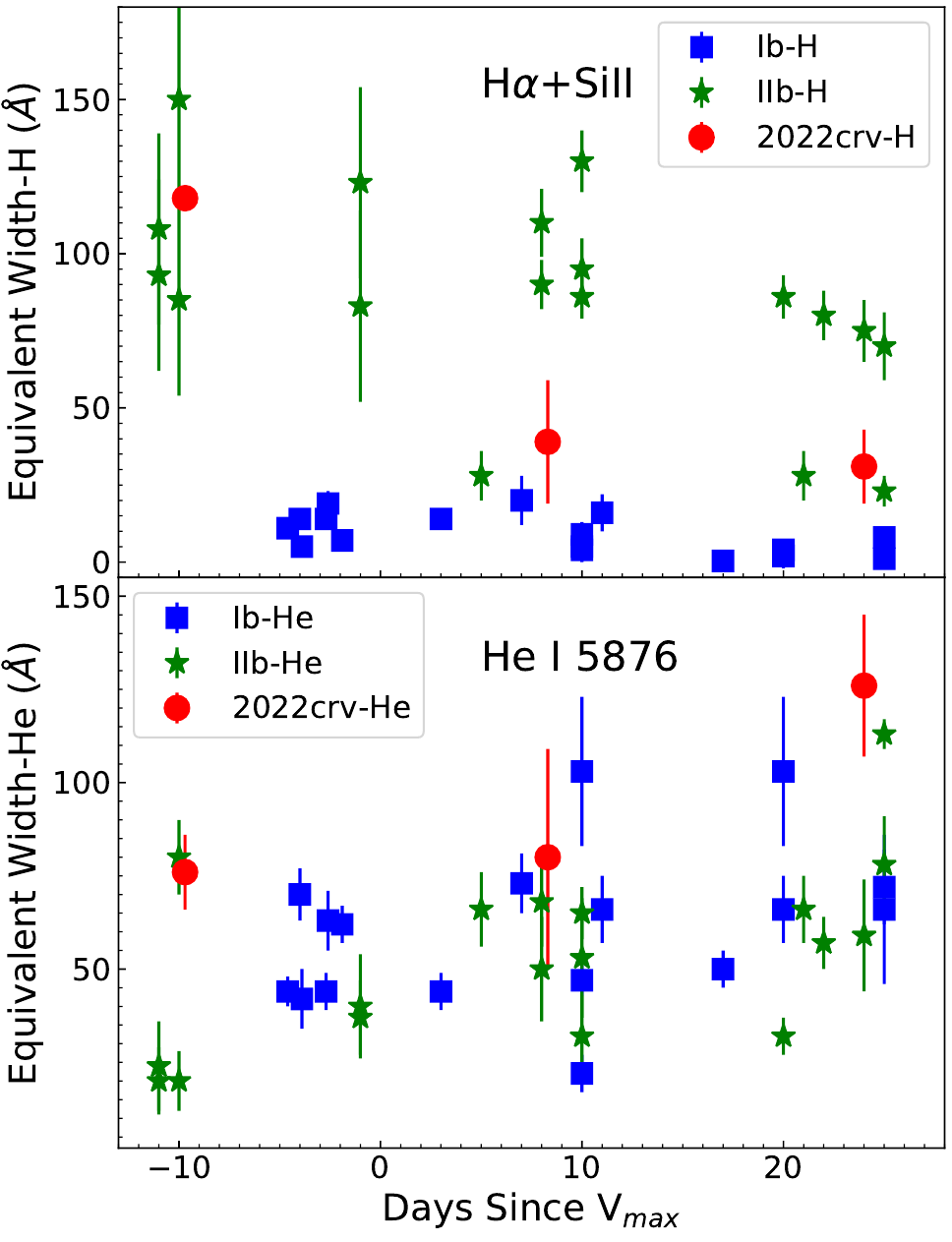}} 
		\caption{The EW evolution of the H$\alpha$+Si\,{\sc ii} (upper) and the He {\sc i} 5876\,\AA\ (lower) lines. For the comparison sample, the EWs of H$\alpha$ and He {\sc i} are calculated for those SNe referenced in Table~\ref{tab:compsample}.}
		\label{fig:eqw}
	\end{figure}

	\begin{figure}
		\centering   
		\resizebox{\hsize}{!}{\includegraphics{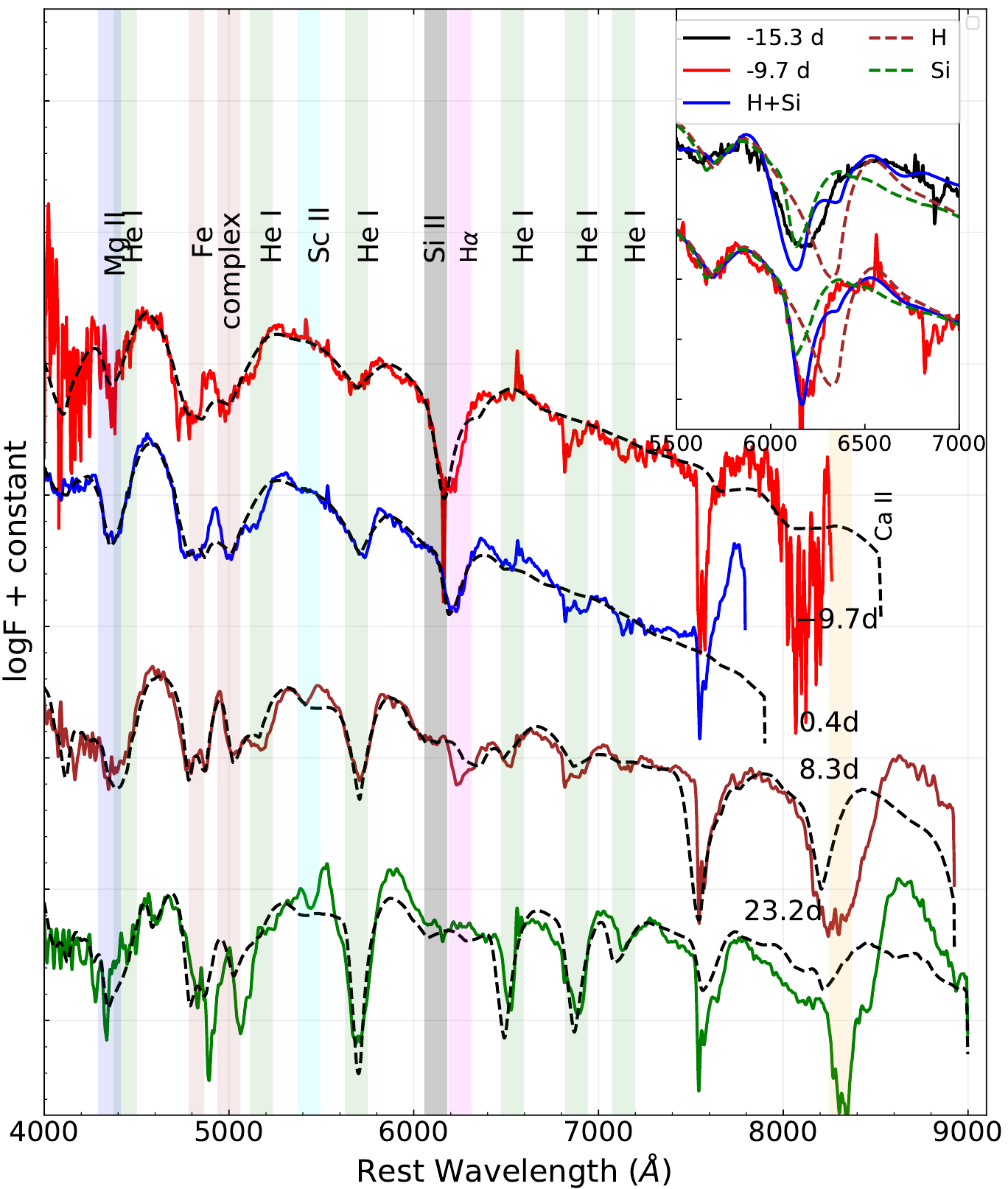}} 
		\caption{\texttt{SYNAPPS} modeling of the early ($-$\,9.7 d) up to a few weeks post-maximum (+23.2 d) spectra of SN~2022crv (shown in black), marked with the lines identified from the spectral modeling. The 6200\,\AA~ dip is reproduced by a combination of H$\alpha$ and Si {\sc ii}, which we can see in the inset showing the spectra at $-$\,15.3 d and $-$\,9.7 d.}
		\label{fig:synapps}
	\end{figure}

	Figure~\ref{fig:precomp} shows the pre-maximum spectral comparison plot of SN~2022crv with other extended SNe~IIb (eSNe~IIb), compact SNe~IIb (cSNe~IIb) and SNe~Ib. The spectral comparison sample is collected from WiseRep \citep{wiserep}\footnote{\url{https://www.wiserep.org/}}, and all the relevant papers are cited in Table~\ref{tab:compsample}. The spectrum of SN~2022crv shows a distinct dip at 6200\,\AA. This feature matches well with those seen in cSNe~IIb, especially SN~2020acat. In addition to the feature at 6200\,\AA, the He {\sc i} features are well-developed in SN~2022crv. This is reminiscent of the behavior seen in SNe IIb, for which the first feature is interpreted as H$\alpha$. Assuming this is also H$\alpha$ in SN~2022crv, the spectrum of SN~2022crv matches with cSNe~IIb in terms of H-richness contrary to the shallow H$\alpha$ seen in eSNe~IIb. In addition, the SNe~Ib in the comparison sample have a prominent dip of He {\sc i} compared to SN~2022crv. The inspection here indicates that SN~2022crv had retained a very thin H envelope, showing its similarity with a cSN~IIb.

	\begin{figure*}
		\centering   
		\resizebox{0.8\hsize}{!}{\includegraphics{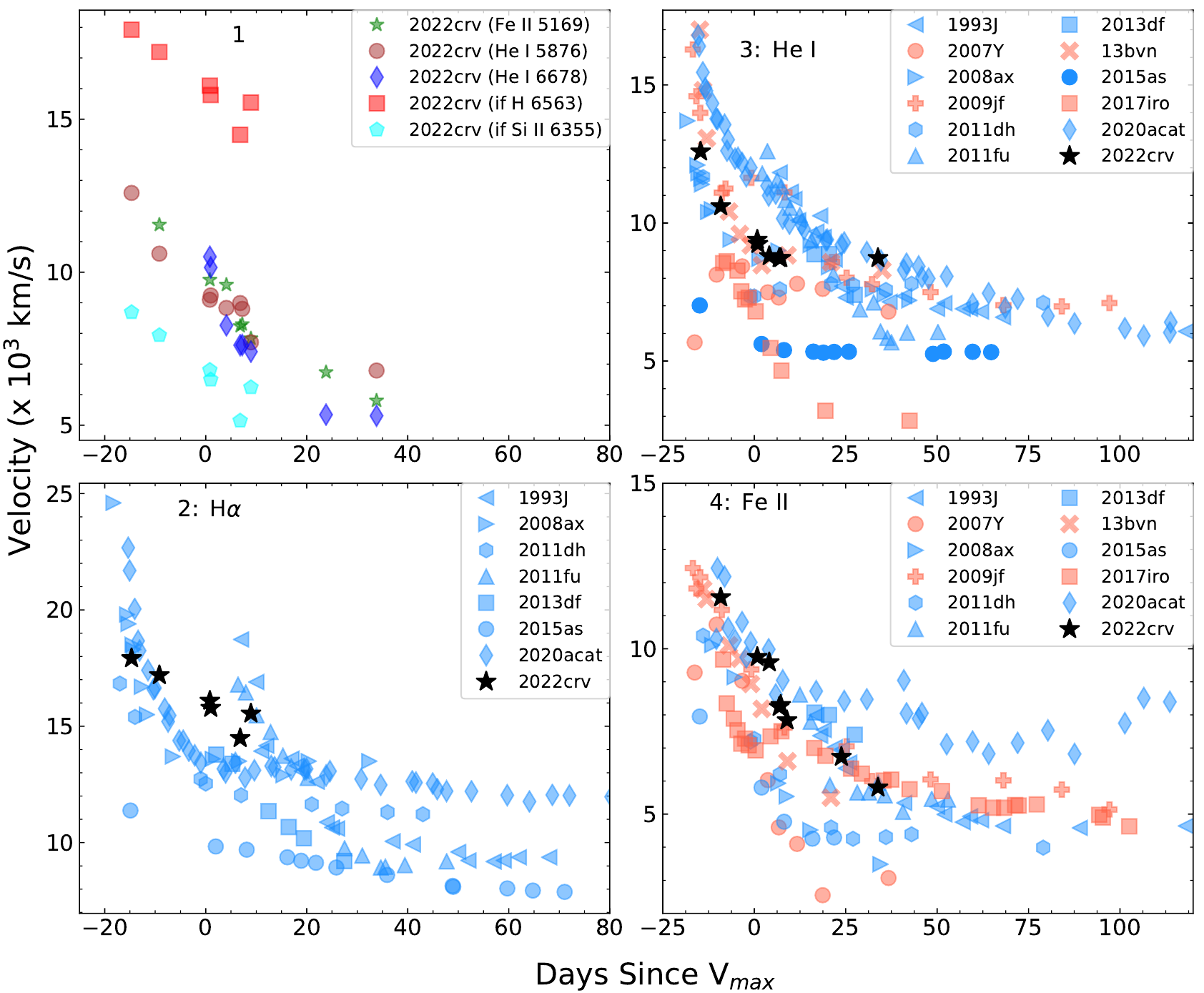}} 
		\caption{The velocity evolution of several lines seen in SN~2022crv. In panels 2-4, they are compared with the SNe~IIb (blue color) and SNe~Ib samples (red color). The error in the line velocity measurements can be as large as 500 km s$^{-1}$, but this is comparable with the symbol sizes used here. The velocities are estimated using the absorption minima of the P-Cygni profiles for SN~2022crv. The velocities for the comparison sample are adopted from references in Table~\ref{tab:compsample}. }
		\label{fig:velevol}
	\end{figure*}

	Figure~\ref{fig:10daycomp} shows the post-maximum spectral comparison plot. The ``W" feature is still visible in SN~2022crv, while the feature in the other SNe~IIb in the comparison sample has vanished completely. SNe~Ib, however, still shows the ``W" features. The He {\sc i} feature at 5876\,\AA\ in SN 2022crv is similar to those seen in SN Ib iPTF13bvn and SN Ib 2015ap at this stage. The interesting observation at this stage is the similarity of the `H$\alpha$' feature of SN~2022crv to SNe~Ib and some of cSNe~IIb, despite its similarity to SNe~IIb in the earlier phase; it has substantially diminished over time in SN 2022crv.  The spectrum of SN 2022crv in this plot extends to a slightly redder portion of the optical window, covering  Ca {\sc ii} NIR features. It is seen that the Ca {\sc ii} NIR features are very strong in SN~2022crv. Indeed, with the caveat that the Ca {\sc ii} NIR features are placed near the edge of the spectra in the other phases, it is possible that Ca {\sc ii} NIR features are always strong in SN~2022crv, and this behavior is similar to SNe~Ib. 
	
	Figure~\ref{fig:25daycomp} shows the late post-maximum spectrum of SN~2022crv as compared to other SE-SNe. SN~2022crv shows very strong He {\sc i} dips. The plot shows that along with the He {\sc i} 5876\,\AA\ feature, strong absorptions are also noticed at 6678\,\AA\ and 7065\,\AA. The overall He strength in SN~2022crv is among the strongest in SNe~IIb and SNe~Ib at this phase.
	
	In summary, the overall spectral evolution of SN~2022crv generally traces that of SN~Ib, e.g., in He features, the `Fe II blend' at 5000\,\AA, and the Ca {\sc ii} NIR features. However, the feature distinct from SNe Ib is seen at $\sim 6200 $\AA\, especially in the pre-maximum phase. The feature is similar to those found in cSNe IIb 2008ax and 2020acat and most likely reflects a substantial contribution by H$\alpha$ in the early phase; this motivates the `SN~IIb' classification for SN 2022crv. Indeed, the above inspections suggest that SN~2022crv is also similar to these cSNe IIb in the spectral evolution, representing a boundary between SNe Ib and cSNe IIb. 
	
	Figure~\ref{fig:eqw} shows the evolution of the H$\alpha$+Si,{\sc ii} and He {\sc i} EWs. We estimated the EWs of these lines using the absorption profiles of the P-Cygni lines, where the flux is normalized to a local continuum. From the upper panel, it is seen that SNe~IIb and SNe~Ib are clearly distinguished. SN~2022crv has a pre-maximum EW similar to other SNe~IIb. With the caveat that data for SNe~Ib in the pre-maximum phase are missing, it indicates that SN~2022crv belongs to the SN~IIb class rather than the SN~Ib, in view of the pre-maximum EW of the H$\alpha$+Si {\sc ii} feature. The EW of the feature in SN~2022crv has decreased substantially by +10 d. This marks the transition of this SN from IIb to Ib. On +25 d, we see that H$\alpha$ has not changed much with respect to the earlier phase and stayed at the level seen in cSNe~IIb and SNe~Ib. 
	
	In the lower panel of Figure~\ref{fig:eqw}, we can see that the EW of the He {\sc i} 5876\,\AA\ line continuously increases in strength as a function of time, for SNe~IIb and Ib in general; the EW of the He {\sc i} 5876\,\AA~ line is practically indistinguishable between these subclasses, and SN~2022crv follows this behavior. 
	By +25 d, we see a prominent rise in the He~{\sc i} EW for SN~2022crv relative to the other SNe~IIb and SNe~Ib in our comparison sample, while it is still within the range expected for SNe~IIb and Ib. This behavior is indeed similar to that seen in cSN~IIb 2008ax. 
	
	In summary, from the evolution of the EWs we conclude that the H$\alpha$+Si {\sc ii} feature initially showed a high EW consistent with those seen in SNe IIb. As time goes by, in the post-maximum phase the feature has weakened substantially, to a level similar to those found in SNe Ib but much weaker than SNe IIb. Meanwhile, the He {\sc i} feature has strengthened over time, consistent with the general behavior seen in SNe IIb and Ib. SN 2022crv thus shows a beautiful transition from the IIb class to Ib.
	
	To strengthen the case for the SN IIb classification through line identification, we used the open-source spectral fitting software \texttt{SYNAPPS} \citep{2013ascl.soft08007T} to reproduce the pre-peak (--9.7 d), peak (+0.4 d) and two post-peak (+8.3 d and +23.2 d) spectra of SN~2022crv (see Figure~\ref{fig:synapps}). \texttt{SYNAPPS} assumes spherical symmetry, homologous expansion, and photosphere emitting blackbody continuum. The emission and absorption lines are formed by resonant scattering, assuming the Sobolev approximation. The velocity at the photosphere, the optical depth of a reference line for each ion, and the minimum and maximum velocities on the distribution of each ion are the fitting parameters for each spectrum. For modeling the SN~2022crv spectra, we used H~{\sc i}, He~{\sc i}, O~{\sc i}, Mg~{\sc ii}, Ca~{\sc ii}, Fe~{\sc ii} and Sc~{\sc ii}. We obtained a maximum velocity of 30,000 km s$^{-1}$ for all the ions, though this value is not strongly constrained. 
	
	In the $-$9.7 d spectrum, a combination of H$\alpha$ and Si {\sc ii} reproduces well the absorption dip at 6200\,\AA. This further justifies that a non-negligible amount of H is present in SN 2022crv. The inset plot of the figure shows that the absorption dip at 6200 \AA~ for the $-$15.3 d and the $-$9.7 d spectra is better reproduced by a combination of H$\alpha$ and Si {\sc ii}, rather than using only one of them. The trace of H$\alpha$ however diminishes rapidly, and He {\sc i} starts dominating the spectrum later on, as seen in the spectrum on +8.3 d.
	
	The photospheric velocities obtained from the \texttt{SYNAPPS} fit to the $-$9.7, +0.4, +8.3, and +23.2 d spectra are consistent with the Fe {\sc ii} velocities we estimate next in Section~\ref{vel}. It is derived to be 9400 km s$^{-1}$ at maximum light. In all the epochs, the \textquoteleft detach' parameter has been deactivated in the fits, meaning that all the ions are distributed (at least) down to the photosphere rather than concentrated at high velocities. 
	
	\subsection{Velocity evolution}
	\label{vel}
	
	Figure~\ref{fig:velevol} shows the velocity evolution of SN~2022crv compared with other SE-SNe. We estimated the velocities of Fe {\sc ii} 5169\,\AA, He {\sc i} 5876\,\AA, and H$\alpha$ 6563\,\AA\ by fitting a Gaussian profile to the absorption trough after correcting the spectra for the redshift of the host galaxy. The velocities of SN~2022crv were compared with the sample of SNe~IIb and Ib collected from the literature in Table~\ref{tab:compsample}. The first panel (1) of Figure~\ref{fig:velevol} shows the velocities of H$\alpha$, Fe {\sc ii} 5169, Si {\sc ii} 6355, He {\sc i} 5876, and He {\sc i} 6678 for SN~2022crv. Assuming that the absorption trough at $\sim$\,6200 \AA\ arises from H$\alpha$, the estimated velocities drop from 18,000 km s$^{-1}$ to 14,500 km s$^{-1}$ between $\rm -15.3$\,d and $\rm +8.3$\,d. If instead this absorption feature is attributed to Si {\sc ii} 6355 \AA, the line velocity of Si would be evolving from 8,600 km s$^{-1}$ to 5,000 km s$^{-1}$, which is lower than for Fe {\sc ii} (Figure~\ref{fig:velevol}); assuming the Fe {\sc ii} traces the photosphere \citep{2005A&A...437..667D}, the behavior suggests that the feature at $\sim 6200$\AA\ could not be created only by Si {\sc ii} and the contribution by H$\alpha$ is likely substantial, as is consistent with the result of the \texttt{SYNAPPS} spectral modeling (Section \ref{spec-comp} and Figure~\ref{fig:synapps}). 
	
	Panel 2 in Figure~\ref{fig:velevol} further supports this view. Assuming the feature is created by H$\alpha$, the velocity evolution is similar to the other SNe~IIb, with one key difference that the H$\alpha$ feature vanishes quite early for SN~2022crv. This behavior indicates that the SN~transitioned early on to a SN~Ib. The H$\alpha$ velocity of SN 2022crv is on the high end among SNe~IIb, possibly due to the blend with Si {\sc ii}, which indicates a thinner H envelope in SN 2022crv than the other SNe IIb.
	
	Panel 3 of Figure~\ref{fig:velevol} shows the velocity evolution of the He {\sc i} 5876 \AA~ line, which decreases from 12,500 km s$^{-1}$ to 7,000 km s$^{-1}$. Panel 4 of Figure~\ref{fig:velevol} shows the velocities of Fe {\sc ii} 5169\,\AA, which decrease from 12,000 km s$^{-1}$ to 6,500 km s$^{-1}$. In general, the velocities of He {\sc i} 5876 \AA\ and Fe {\sc ii} 5169 \AA\ (which trances the photosphere) are within the diversity seen for SNe IIb and Ib, suggesting that the key difference between SN 2022crv and SNe IIb, as well as that between SN 2022crv and SN Ib, is mainly in the nature of the H-rich envelope. 
	
	To summarise, the velocity evolution supports the idea that the absorption trough around 6200\,\AA\ is generated by a combination of H$\alpha$ and Si {\sc ii}. The He {\sc i} and Fe {\sc ii} velocities are similar to most SNe~Ib. The H$\alpha$ + Si {\sc ii} and He {\sc i} velocities thus support the metamorphosis of SN~2022crv from  SN~IIb to SN~Ib.
	
	\begin{figure}
		\centering   
		\resizebox{\hsize}{!}{\includegraphics{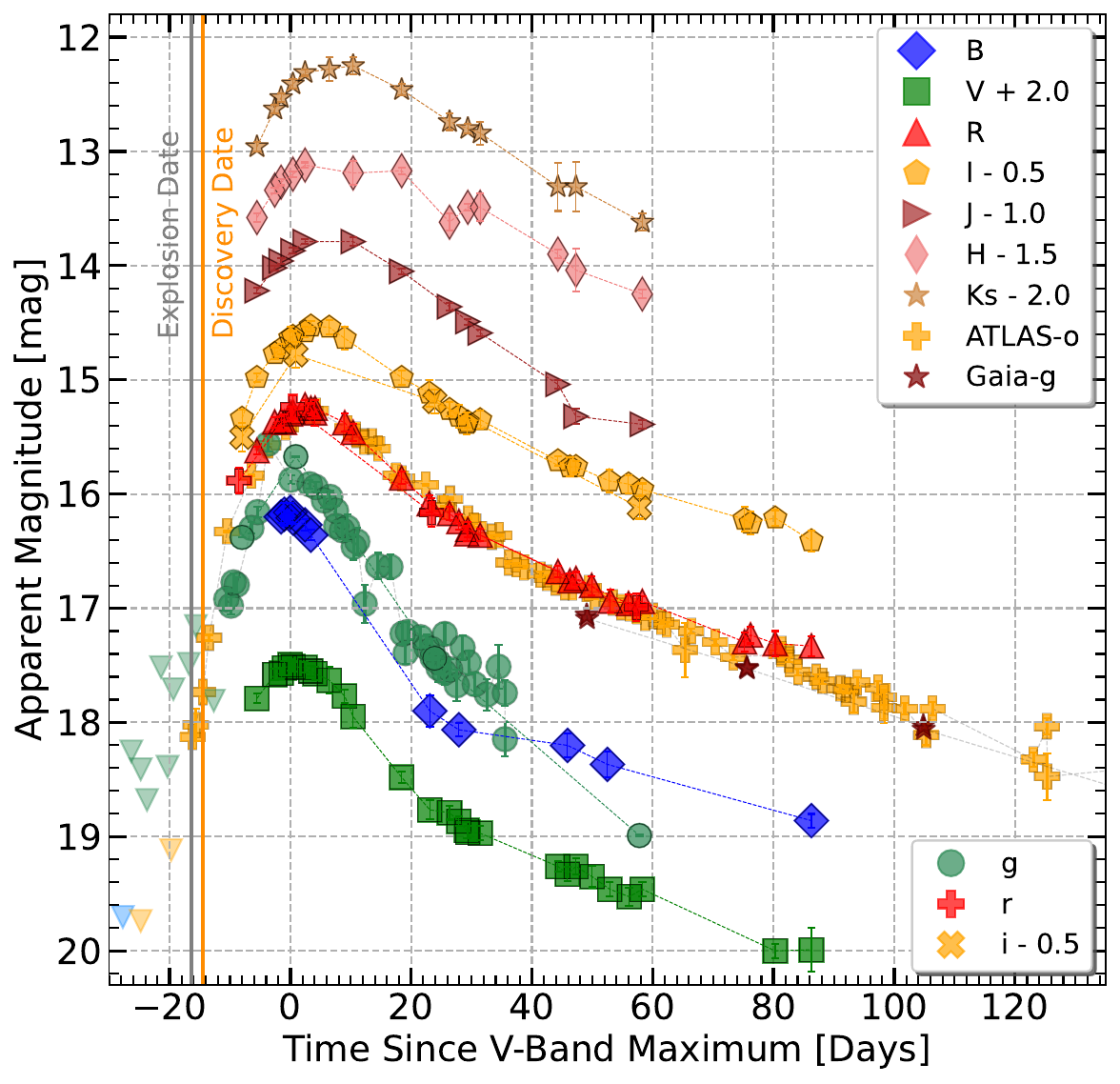}} 
		\caption{The apparent magnitude light curves of SN~2022crv in the {\it BgVRIJHK} filters. The $g$-band is adapted from Seimei-TriCCS and ASASSN.}
		\label{fig:applc}
	\end{figure}
	
	\section{Photometric Evolution}
	\label{phot}
	
	\begin{figure}
		\centering   
		\resizebox{\hsize}{!}{\includegraphics{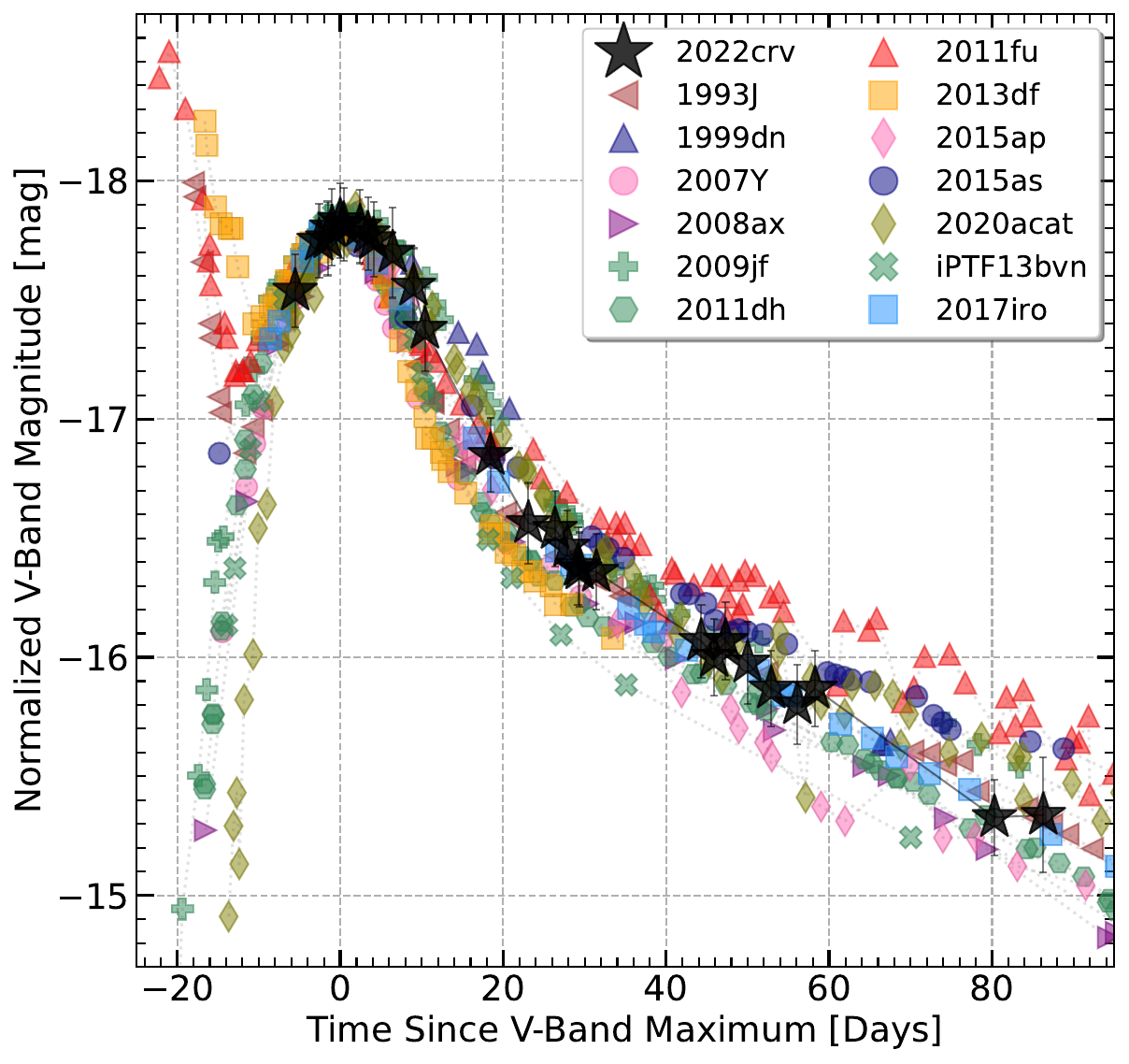}} 
		\caption{The $V$-band absolute magnitude light curve of SN~2022crv. For the comparison SNe~IIb and Ib, the magnitude scale is normalized to the peak magnitude of SN 2022crv. The values of the reference sample are taken from the papers cited in Table~\ref{tab:compsample}.}
		\label{fig:vbandcomplc}
	\end{figure}
	
	\begin{figure}
		\centering   
		\resizebox{\hsize}{!}{\includegraphics{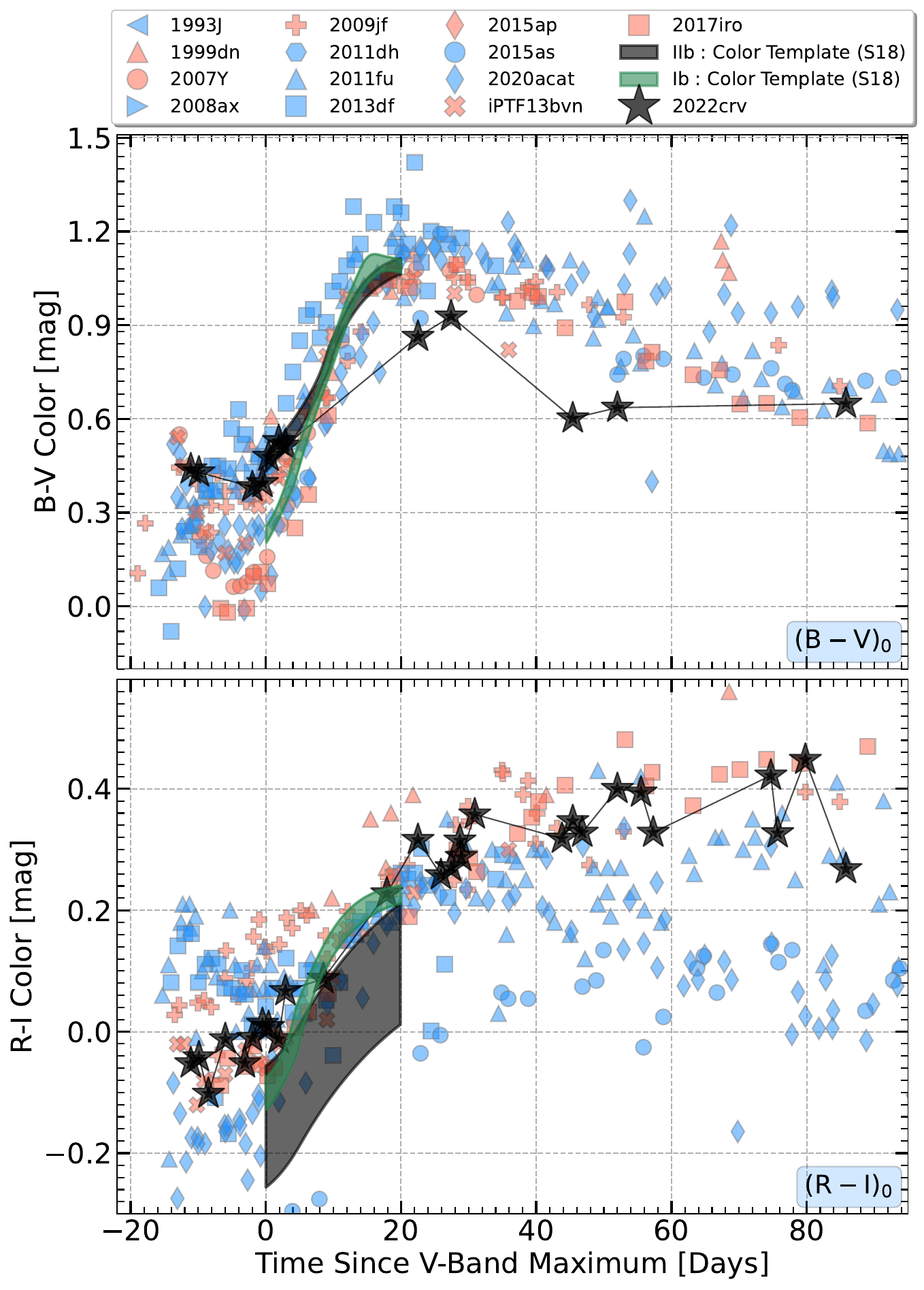}} 
		\caption{Optical color evolution of SN~2022crv compared to a sample of SE-SNe. The SE-SNe sample has been corrected for Milky Way and host galaxy extinction. The templates from \citet{2018A&A...609A.135S} are also overplotted. The blue-colored points are for SNe~IIb, and the red-colored points are for SNe~Ib. The data for the comparison sample are taken from the papers cited in Table~\ref{tab:compsample}.}
		\label{fig:colorevol}
	\end{figure}
	
	\begin{figure}
		\centering   
		\resizebox{\hsize}{!}{\includegraphics{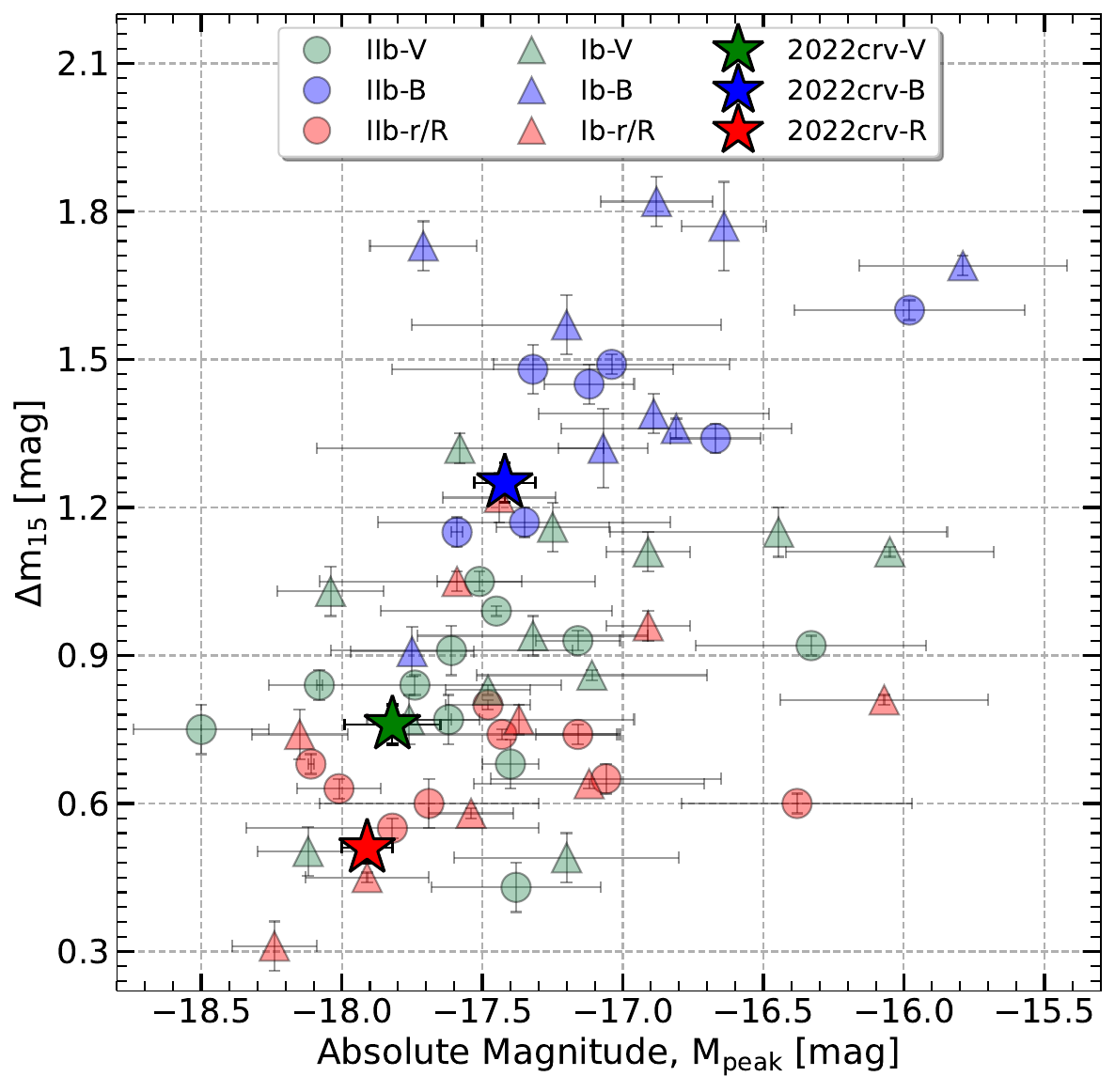}} 
		\caption{Peak absolute magnitude versus $\Delta$m$_{15}$ of SN~2022crv in comparison with a sample of SNe~IIb and SNe~Ib from \citet{2018AA...609A.136T} and Table~\ref{tab:compsample}.}
		\label{fig:deltam15}
	\end{figure}
	
	\begin{table*}
\centering
\caption{Properties of the comparison sample}
\label{tab:compsample}

\smallskip
%\footnote
\begin{tabular}{c c c c c c c c c}
\hline \hline
                SNe        & Host galaxy & Distance   & Extinction & SN Type    & Absmag    & $\Delta$m$_{15}$ & Decay rate$^\ddagger$ &  Reference$^\dagger$  \\
                                  &             & (Mpc)      & E(B-V)      &            & $V$-band  & $V$-band   &  mag/100day  &  \\
\hline
                        SN 1993J  & M81    & 3.63$\pm$0.05       & 0.18 & IIb  & -17.59$\pm$0.13 & 1.65   &  1.73 & 1,2      \\
                        SN 2008ax  & NGC4490    & 9.47$\pm$1.3       & 0.30 & IIb   & -17.61$\pm$0.43  & 0.91   & 1.90 & 3,4,17  \\
                        SN 2011dh & M51    & 9.06$\pm$0.63       & 0.035 & IIb   & -17.12$\pm$0.18 & 0.98 & 1.76  & 5,17     \\
                        SN 2011fu  & UGC1626    & 74.54$\pm$5.22       & 0.22 & IIb  & -18.50$\pm$0.24 & 1.75 & 1.78 & 6,17      \\
                        SN 2013df  & NGC4414    & 8.86$\pm$0.62       & 0.10 & IIb   & -16.85$\pm$0.08 &  0.43 & 1.81 & 7,17  \\
                        SN 2015as & UGC5460    & 20.43$\pm$1.43       & 0.008 & IIb   & -16.82$\pm$0.18 & 0.68 & 1.85 & 8,17     \\
                        SN 2020acat &  PGC037027         & 35.3$\pm$4.4         & 0.021 & IIb    & -17.62$\pm$0.11      & 0.77 & 1.67 & 9,17   \\
\hline\hline
                        SN 1999dn  & NGC7714 &  38.5$\pm$2.31  & 0.10 &  Ib  & -17.20$\pm$0.40   &  0.49 & 1.54 &  10,14,17 \\
                        SN 2007Y   & NGC1187 &  18.14$\pm$1.27 & 0.112 & Ib & -16.47$\pm$0.60 & 1.15  & 1.73 & 11,14 \\
                        SN 2009jf  & NGC7479 &  33.7$\pm$1.68       & 0.11  & Ib & -17.96$\pm$0.19  & 0.50 & 1.36 & 12,14 \\
                        iPTF13bvn  & NGC5806 &  23.71$\pm$1.66   &  0.17 & Ib & -17.23$\pm$0.20 & 1.16 & --  & 13 \\     
                        SN 2015ap  & IC1776  &  45.10$\pm$3.16   & 0.037 &  Ib & -18.04$\pm$0.19 & 1.03 & 1.55  & 14 \\
                        SN 2017iro & NGC5480 &  33.64$\pm$2.36   & 0.28  &  Ib & -17.76$\pm$0.15 & 0.77 & 1.73 & 15 \\
\hline\hline
                    SNe~IIb     & $--$ & $--$ & $--$ & $--$ & -17.40$\pm$0.55 &  0.93$\pm$0.08 & $--$ & 16 \\  
                    SNe~Ib       & $--$ & $--$ & $--$ & $--$ & -17.07$\pm$0.56 & 1.03$\pm$0.19 & $--$ & 16\\
\hline
\end{tabular}
\newline
$^\dagger$ REFERENCES.-- (1)\cite{1995A&AS..110..513B}, This work; (2) \cite{2004Natur.427..129M}, This work; (3)\cite{2008MNRAS.389..955P};  (4)\cite{2011MNRAS.413.2140T}; (5)\cite{2013MNRAS.433....2S}; (6)\cite{2013MNRAS.431..308K}; (7)\cite{2014MNRAS.445.1647M}; (8)\cite{2018MNRAS.476.3611G}; (9)\cite{2022MNRAS.513.5540M}; (10)\cite{2011MNRAS.411.2726B}; (11)\cite{2009ApJ...696..713S}; (12)\cite{2011MNRAS.416.3138V}; (13)\cite{2014MNRAS.445.1932S}; (14)\cite{2020MNRAS.497.3770G}; (15) \cite{2022ApJ...927...61K}; (16)\cite{2018AA...609A.136T}; (17) NED
$^\ddagger$ The $V$-band decay rate in the phase between 40 and $>100$ days. \\
%\label{tab:photometric_parameters_different_SNe}.
\end{table*}

	The multi-band light curve evolution of SN~2022crv is shown in Figure~\ref{fig:applc}. The $V$-band light curve is compared to other SE-SNe in Figure \ref{fig:vbandcomplc}, where the absolute magnitude values of the comparison SNe are shifted in the magnitude scale to match that of SN 2022crv. We estimated all the light curve parameters of SN 2022crv by fitting the data to an analytical formulation by \cite{2018AA...609A.136T}, which is a modified version of \cite{1996ApJ...471L..37V} to apply to SE-SNe. The shape of the SE-SN light curves can be represented by three components; (i) an initial exponential rise, (ii) a Gaussian-like peak, and (iii) a late linear decay. The maxima of the light curves and the other parameters obtained from this fitting are tabulated in Table \ref{tab:params}. We could trace the maxima in all the filters thanks to the pre-maximum discovery. The $V$-band light curve had a rise-time of $\sim$15 d. A lag of $\sim$7 d in the rise time is robustly derived between the $B$ and the $I$ band. Most of the SE-SNe peak earlier in bluer bands, and maxima in other bands follow owing to cooling of the SN~ejecta, a trend which is also noticed for SN~2022crv. 
	
	\begin{table*}
\centering
\caption{Observed parameters of SN~2022crv}
\label{tab:params}
\smallskip
%\footnote
\begin{tabular}{l c c c c c}
\hline \hline
2022crv                                       & $B$ band           & $V$ band                      & $R$ band              & $I$ band   \\
\hline
JD (maximum)                                & 2459643.60$\pm$0.10     & 2459644.80$\pm$0.10       & 2459647.10$\pm$0.10       & 2459650.30$\pm$0.10     \\
Magnitude at maximum (mag)                   & 16.16$\pm$0.01      & 15.53$\pm$0.01         & 15.28$\pm$0.01        & 15.06$\pm$0.01 \\
Absolute magnitude at maximum (mag)          & -17.42$\pm$0.12      & -17.82$\pm$0.17       & -17.91$\pm$0.09      & -17.94$\pm$0.13    \\
$\Delta$m$_{15}$ (mag)                       & 1.25$\pm$0.04      & 0.76$\pm$0.04       &   0.51$\pm$0.03     & 0.49$\pm$0.02  \\
$\Delta$m$_{40}$ (mag)                       & 2.02$\pm$0.01     &  1.69$\pm$0.01    &  1.42$\pm$0.01   &   1.15$\pm$0.01   \\
Rise Time (day)                              &    13.9                &  15.1                     & 17.4                &  20.6               \\
% \hline
% Colour at maximum$^\dagger$                          &          &      &    &   	       \\
% $(B-V)_{0}$                                          & 0.91   &      &    &   	       \\
% $(V-R)_{0}$                                          & 0.60   &      &    &   	       \\
% $(V-I)_{0}$                                          & 0.42   &      &    &   	       \\
% $(R-I)_{0}$                                          & 0.30   &      &    &   	       \\
\hline
Decline rate mag (100 days)$^{-1}$   & 1.36$\pm$0.26     & 1.55$\pm$0.31    & 2.60$\pm$0.26   & 1.87$\pm$0.08 \\
Time range (40 - 86 d)                   &            &          &           &  \\
\hline \hline
2022crv                                        & $J$ band     & $H$ band                      & $K$ band              &   \\
\hline
JD (maximum)                                & 2459651.40$\pm$0.10      & 2459653.9$\pm$0.10       &  2459654.62$\pm$0.11        &       \\
Magnitude at maximum (mag)                   & 14.78$\pm$0.01       & 14.63$\pm$0.01        & 14.28$\pm$0.01         &     \\
Absolute magnitude at maximum (mag)          & -18.18$\pm$0.16      & -18.24$\pm$0.17       & -18.52$\pm$0.19       &     \\
$\Delta$m$_{15}$ (mag)                       &  0.37$\pm$0.02     & 1.90$\pm$0.06      & 2.25$\pm$0.01       & \\
$\Delta$m$_{40}$ (mag)                       &  1.46$\pm$0.01     &  2.71$\pm$0.40                     &    2.37$\pm$0.34                   & \\
Rise Time                                    &   21.7              & 24.2     &  24.92  &  \\
\hline
Decline rate mag (100 days)$^{-1}$     & 4.01$\pm$0.17     & 2.92$\pm$0.08    & 2.42$\pm$0.35   \\
Time range (46 - 102 d)                   &            &          &           &  \\
\hline
\hline
\end{tabular}
\newline
\end{table*}

	The $\Delta$m$_{15}(V)$ estimated from the light curve of SN~2022crv is 0.76$\pm$0.04 (see Table~\ref{tab:params}). The $\Delta$m$_{15}(V)$ of SN~2022crv is lower than the average $\Delta$m$_{15}(V)$ quoted by \cite{2018AA...609A.136T} indicating the slow-evolving nature of the SN. The relatively slow evolution of SN 2022crv around the peak is also seen in other bands see below.

	\cite{2018AA...609A.136T} found that the late-time decay rates (from $\sim 40 - 100$ d) of a sample of SNe~IIb and Ib are 1.6\,--\,2.1 mag / 100 d and 1.4\,--\,1.8 mag / 100 d, respectively, in the $V$ band (see also Table \ref{tab:compsample}). As shown in Figure \ref{fig:vbandcomplc}, the late-time decay rate of SN 2022crv is consistent with other SNe IIb and Ib. The limited data for SN 2022crv results in a relatively large error in the decay rate (Table~\ref{tab:compsample}), and thus it is not clear if the slow evolution as compared to other SE-SNe seen around the peak persists in the late phase or not. In any case, the decay rate is higher than the $^{56}$Co $\,\to\,$ $^{56}$Fe decay rate, which corresponds to events with higher gamma-ray escape fractions due to higher explosion energy to ejecta mass ratios \citep{2022ApJ...927...61K}; this is a typical behavior seen in SE-SNe \citep{maeda2003}. To summarise the light curve parameters, we see that SN~2022crv shows a relatively slow evolution up to +40 d post maximum, but overall it is typical of SE-SNe. 
	
	The evolution in {\it (B-V)$_{0}$} and {\it (R-I)$_{0}$} colors for SN~2022crv are plotted with other members of the comparison sample in Figure~\ref{fig:colorevol}. All the colors are corrected for the extinction values tabulated in Table~\ref{tab:compsample} (see Section~\ref{dist-extinction}). The {\it (B-V)$_{0}$} color evolution of SN~2022crv shows an early red-to-blue transition before the $V$-band maximum. This behavior is shared with single-peaked cSNe~IIb like SNe 2008ax and 2010as, but not with double-peaked SNe (or eSNe~IIb) like SNe 1993J, 2011fu, and 2013df \citep{2016PhDT.......113M}. Up to about +30 d, the color curves become redder again, indicating cooling of the photosphere, and become flatter until $\sim$ +90 d. A similar trend is also noticed in the {\it (R-I)$_{0}$} color. The post-maximum color evolution of SN~2022crv matches reasonably well with the average color evolution templates compiled by \cite{2018A&A...609A.135S} for SE-SNe denoted by shaded regions in the plots, and it is similar to cSNe IIb and SNe Ib in the pre-maximum phase.
	
	\begin{figure}
		\centering   
		\resizebox{\hsize}{!}{\includegraphics{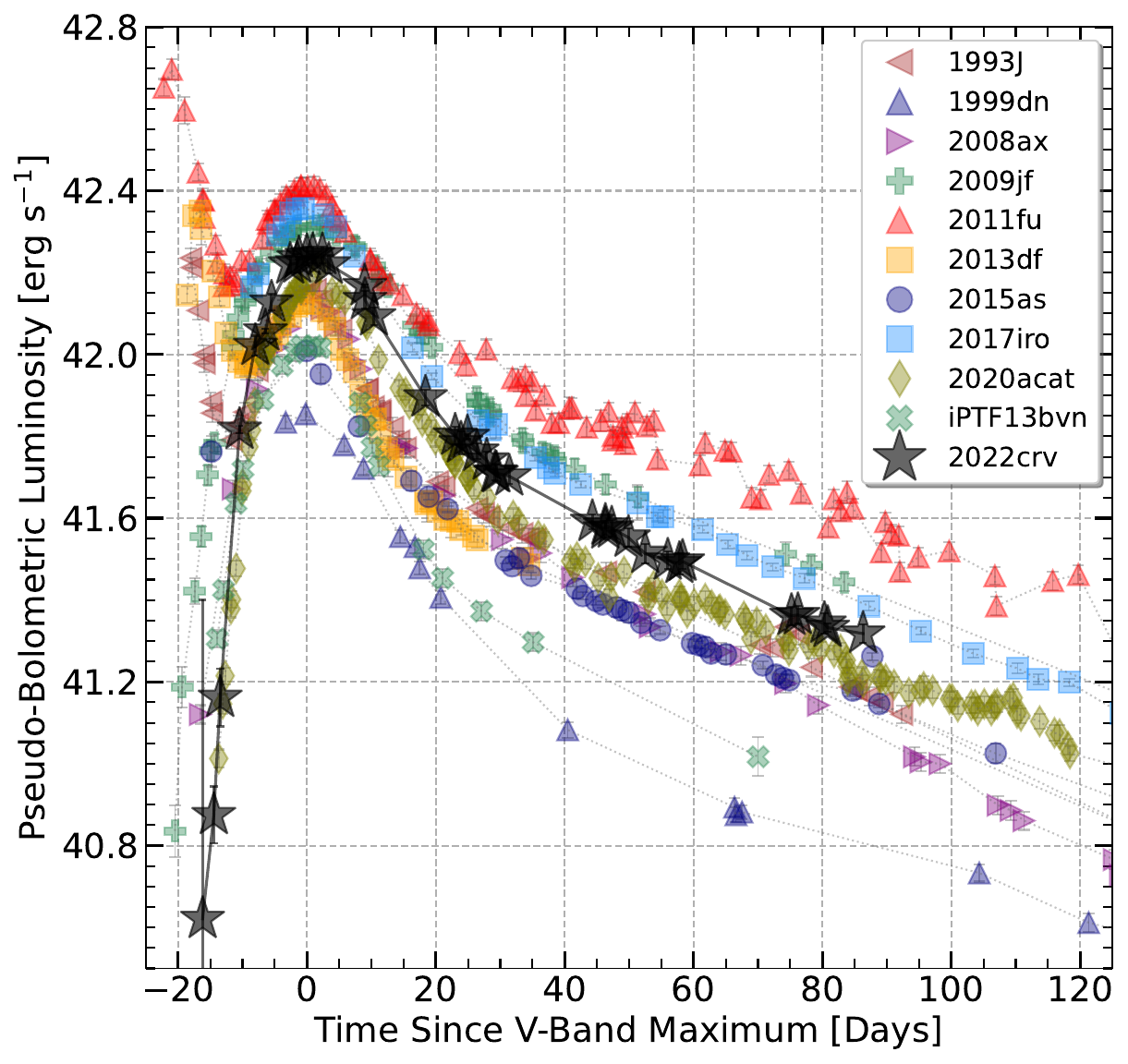}} 
		\caption{The {\it BVRI} bolometric light curve of SN~2022crv generated using {\it SuperBol}. The bolometric light curves of all the comparison objects are also calculated using {\it SuperBol} using the values computed in Table~\ref{tab:compsample}.}
		\label{fig:bollc}
	\end{figure}
	
	\begin{figure}
		\centering   
		\resizebox{\hsize}{!}{\includegraphics{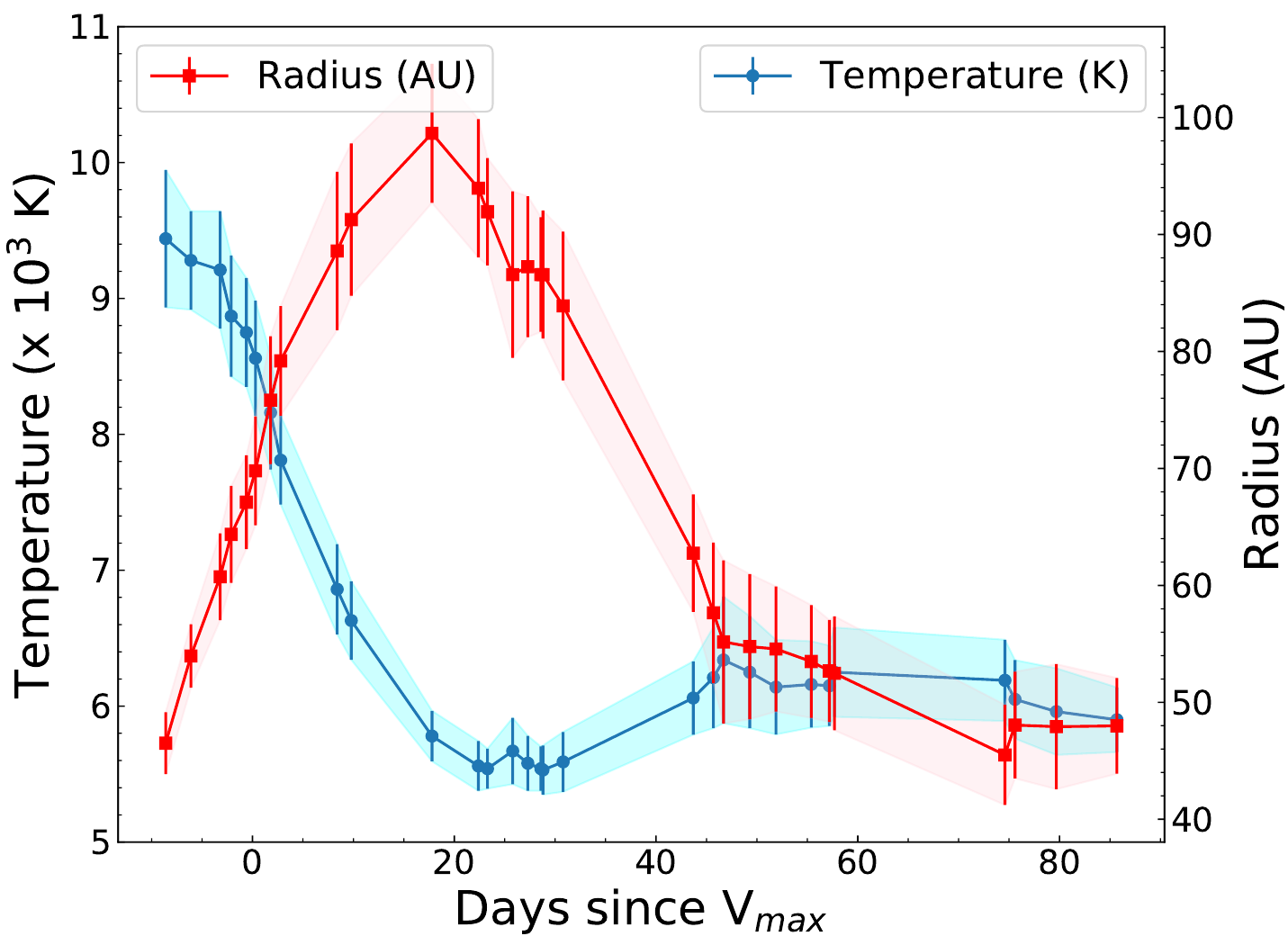}} 
		\caption{The radius and temperature evolution of SN~2022crv up to +\,86\,d. The radii and temperatures are estimated under the blackbody approximation.}
		\label{fig:bol_rad_temp}
	\end{figure}
	
	The $V$-band absolute magnitude of the SE-SN group lies between $-$16.5 mag and $-$19.5 mag \citep{2006AJ....131.2233R, 2011ApJ...741...97D, 2018AA...609A.136T}. The peak M$_{V}$ for SN~2022crv is estimated to be $-$17.82$\pm$0.17 mag. It is consistent with the average value of SNe~IIb ($-$17.40$\pm$0.55) and brighter than the average SNe~Ib ($-$17.07$\pm$0.56) \citep{2018AA...609A.136T}. The peak {\it J} and {\it H} band magnitudes (see Table \ref{tab:params}) of SN~2022crv are brighter than the average NIR magnitudes quoted by \cite{2018AA...609A.136T}. Figure \ref{fig:vbandcomplc} shows that the light curve shape of SN 2022crv is typical of cSNe IIb and SNe Ib. 
	
	Figure~\ref{fig:deltam15} compares the absolute magnitudes of SN~2022crv in different bands with the $\Delta$m$_{15}$ values, as compared with the statistical sample of \cite{2018AA...609A.136T}. No strong correlation is seen between these parameters. While SN 2022crv is one of the slowly evolving members among SE-SNe, its properties lie within the scatter in all the bands; it indicates that the light curve properties, and thus the core properties, are typical of SE-SNe. 
	
	To summarise, SN~2022crv shows a color evolution consistent with cSNe~IIb, indicating the absence of a primary peak. The SN is of average brightness and its light-curve properties are largely consistent with SE-SNe. 
	
	\begin{table*}
\caption{Best fit parameters derived from the bolometric light curve modelling of Nagy and Vinko 2014,2016.}
\centering
\label{tab:NagyVinkomodel}        
\smallskip
%\footnote
\begin{tabular}{c c c c}
\hline \hline
                Parameter        & Core(He-rich) & Shell(H-rich)   & Remarks  \\
                                 & (K=0.06 cm$^{2}$ g$^{-1}$) & (K=0.24 cm$^{2}$ g$^{-1}$) &                  \\
\hline
                         R$_{0}$(cm) &  0.2 x 10$^{11}$ & 0.04 x 10$^{12}$ - 0.29 x 10$^{12}$ & Initial radius of the ejecta   \\
                         T$_{rec}(K)$   & 5500  &  --       &  Recombination Temperature   \\
                         M$_{ej}$(M$_{\odot}$)    & 3.9 & 0.015 - 0.05 & Ejecta mass   \\
                         M$_{Ni}$(M$_{\odot}$)  &   0.112 & 0.0 & Nickel mass    \\
                         E$_{Th}$(foe)  &                0.7    &  0.05  & Thermal energy \\
                         E$_{k}$(foe) &                3.4    &   0.30  & Kinetic energy \\
 \hline \hline                        
\end{tabular}
%\label{tab:photometric_parameters_different_SNe}.
\end{table*}

	\section {Bolometric light curve modeling and estimation of physical parameters}
	\label{bol}
	
	The quasi bolometric light curve of SN~2022crv was constructed using the Python-based code {\sc SuperBol} \citep{2018RNAAS...2..230N}. The {\it BVRI} magnitudes were corrected for extinction as given in Section~\ref{dist-extinction}. The blue bands were extrapolated in the late phases using a constant color, as is derived from the multi-band data on an epoch in which such data are available. The flux integration was performed over the optical wavelengths, and the resultant quasi-bolometric light curve of SN~2022crv is plotted with other SNe~IIb and Ib in Figure~\ref{fig:bollc}. The very early points of SN~2022crv were generated by converting the ATLAS o-band points to bolometric luminosity values with the Python package \texttt{SYNPHOT}, using the SEDs of SN 2020acat as templates given its similarity to SN 2022crv. The figure shows that the bolometric luminosity of SN~2022crv is typical of SE-SNe, with a close similarity to cSN IIb 2020acat. 
	
The blackbody fits for radius and temperatures using {\it BgVRIJHK} magnitudes are plotted in Figure~\ref{fig:bol_rad_temp}. The temperature of the photosphere decreased from 9500 K to 5500 K from -\,9\,d to +\,18\,d, indicating the cooling of the SN ejecta. During the same period, the radius of the outer envelope increased by 50 AU from 45 AU to 95 AU. On the other hand, using our spectroscopic measurements (an average photospheric velocity of $\sim 8,000$ km s$^{-1}$ during the period in consideration), the radius must have been increased by $\sim$\,130 AU during the $\sim$\,27\,d time interval. The spectroscopic and photometric indicators therefore agree within a factor of three, but the difference may be non-negligible. We suspect this might be due to underestimated bolometric luminosity and/or the use of a constant value of photospheric velocity, which is actually decelerating.
	
	The peak properties are mainly determined by the radioactive $^{56}$Ni synthesized in the explosion, the ejecta mass M$_{ej}$, and the kinetic energy E$_{k}$ of the ejecta. We modeled the early photospheric phase of SN~2022crv using the formulation by \cite{2008ApJ...673L.155V} which is based on the original formulation by \cite{1982ApJ...253..785A}. The major assumptions are spherical symmetry, homologous expansion, and a constant opacity ($\kappa$$_{\rm opt}$). The free parameters are M$_{Ni}$ (affecting the peak luminosity) and the diffusion time scale $\tau_{m}$ (controlling the width of the bolometric light curve). Assuming uniform density, the ejecta kinetic energy E$_{\rm k}$ and $\tau_{\rm m}$ are related as:
	
	\begin{equation}\label{eq_1}
		\tau_{\rm m} = \sqrt{2} \left( \frac{\kappa_{\rm opt}}{\beta c} \right)^{1/2} \left( \frac{M_{\rm ej}}{v_{\rm ph}} \right)^{1/2},
	\end{equation}
	
	\begin{equation}\label{eq_2}
		E_{\rm k} \approx \frac{3}{5}\frac{M_{\rm ej}v^2_{\rm ph}}{2},
	\end{equation}
	
	\begin{figure}
		\centering   
		\resizebox{\hsize}{!}{\includegraphics{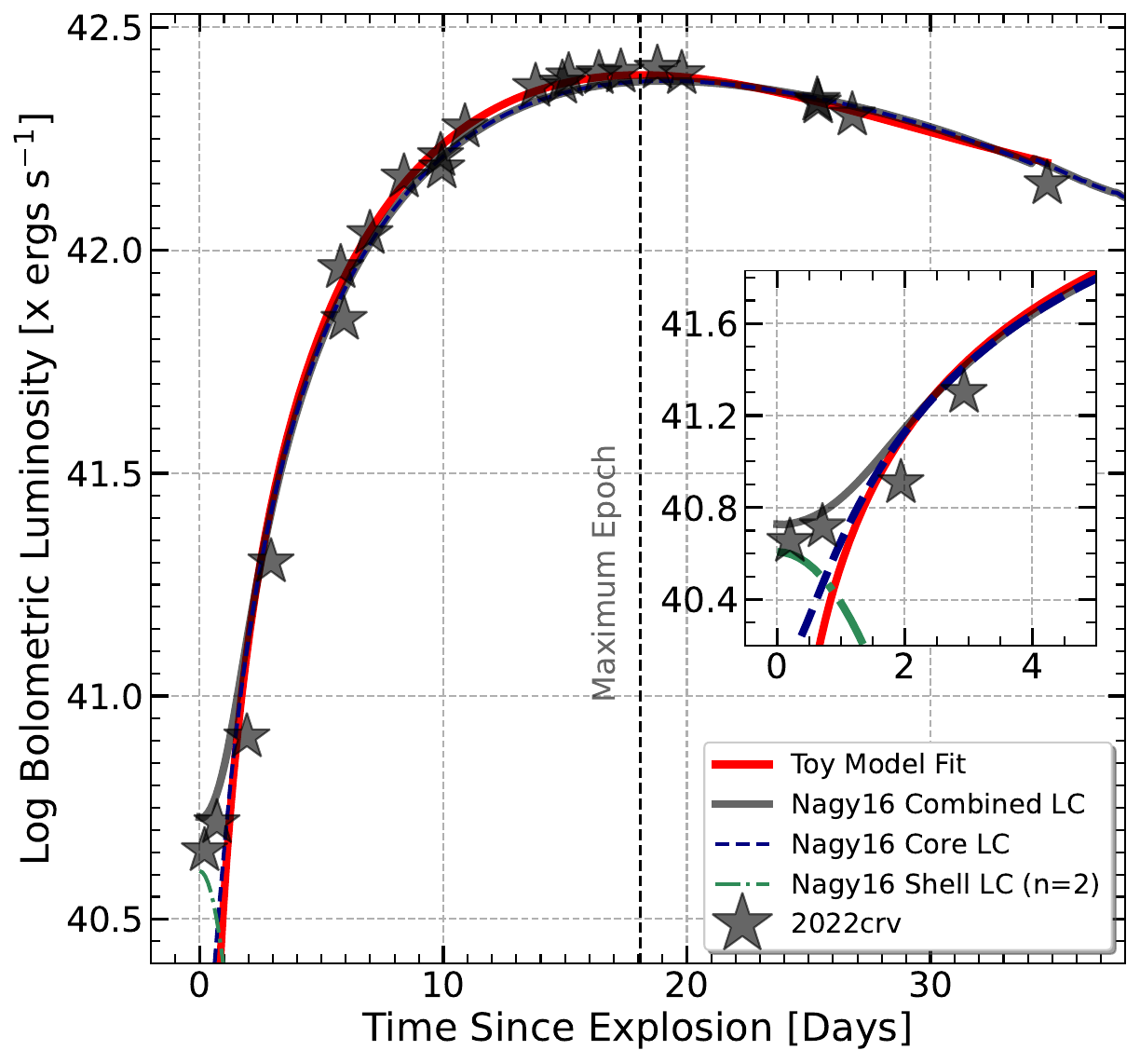}} 
		\caption{The bolometric light curve of SN~2022crv is plotted along with the fit with the toy model of \citet{2008MNRAS.383.1485V}. The best fit from the two-component semi-analytical model proposed by \citet{2016A&A...589A..53N} is also shown along with the contribution from the individual core and shell components.}
		\label{fig:valentifit}
	\end{figure}
	
	\noindent
	where $\beta \approx 13.8$ is a constant of integration \citep{1982ApJ...253..785A} and {\it c} is the speed of light. The optical opacity $\kappa_{\rm opt}$ is adapted to be 0.07 cm$^2$ g$^{-1}$ as is frequently adopted for SE-SNe \citep[e.g.][]{2000AstL...26..797C, 2018AA...609A.136T}. The light curve of SN~2022crv was fitted with this analytical form using least-square optimization up to +36 d post-explosion (Figure \ref{fig:valentifit}). The $^{56}$Ni mass thus obtained is $M_{\rm Ni}$\,=\,0.126$\pm$0.021 M$_{\odot}$ and the diffusion time is $\tau_{\rm m}$\,=\, 16.10$\pm$0.5 d. The ejecta mass and kinetic energy thus obtained are  M$_{\rm ej}$\,=\,3.19 M$_{\odot}$ and E$_{\rm k}$\,=\,1.72$\times 10^{51}$~erg, respectively. 
	
	With the probable existence of the H-rich envelope attached to SN 2022crv, it is highly interesting to constrain the nature of the envelope using the earliest photometric points. The first point, as reported by ATLAS, indeed shows a hint of an excessive emission, which might signal the early envelope-cooling emission (Figures \ref{fig:applc} and \ref{fig:valentifit}). 
	
	To constrain the nature of the H-rich envelope, we used the semi-analytical models by \citet{2014A&A...571A..77N} and \citet{2016A&A...589A..53N}. Here, the bolometric light curve is modeled using a two-component ejecta configuration: an extended, low-mass, H-rich outer envelope and a compact He-rich core. The light curve is thus the combination of radiation from the shock-heated H-ejecta and the radioactive decay of $^{56}$Ni to $^{56}$Co. We adopt $\kappa_{\rm opt}$\,=\,0.24 cm$^{2}$ g$^{-1}$ for the outer layer \citep{1989ApJ...340..396A} and $\kappa_{\rm opt}$\,=\,0.06 cm$^{2}$ g$^{-1}$ for the core. As the photon diffusion time scale is much shorter in the outer shell than in the core, the contributions of the two regions to the overall light curve are well separated. In practice, we first fitted the core properties and then constrained the envelope properties since the strength of the early-cooling emission depends on the underlying $^{56}$Ni-heating light curve. 
	
	The best-fit values of $^{56}$Ni, M$_{ej}$ (core and shell), E$_{k}$ (kinetic energy), E$_{\rm Th}$ (thermal energy), and the radii of the core and shell are given in Table \ref{tab:NagyVinkomodel}. Given the uncertainties in the first ATLAS point and the underlying $^{56}$Ni-heating contribution, we regard the envelope radius obtained here as an upper limit for a given envelope mass (see Section \ref{progenitor} for further details). We find that the constraints of R $< 1 - 3$ R$_{\odot}$ and M $\sim$ 0.015--0.05 M$_{\odot}$ can be placed for the properties of the H-rich envelope of SN 2022crv. The detailed interpretation of the progenitor compactness and its correlation with the theoretical models is given in Section~\ref{progenitor}. The derived core parameters obtained by this analysis \citep{2016A&A...589A..53N} are largely consistent with those derived by the ones obtained with the model of \citet{2008MNRAS.383.1485V}. This is not surprising as the assumptions are similar between the two formulations. The kinetic energy is overestimated in the latter model, but it likely stems from some difference in details about how the diffusion time scale is converted to the mass and kinetic energy. 
	
	Combining all the above analyses of SN~2022crv, we find that the $^{56}$Ni mass is M$_{\rm Ni}$\,=\,0.12$\pm$0.05 M$_{\odot}$ and ejecta mass is in the range of 3.2--3.9 M$_{\odot}$. The ejecta parameters of SN~2022crv are consistent with the range found for SE-SNe but with the ejecta mass on the higher side (Table \ref{tab:compbolsample}). 
	
	\begin{table}
\caption{Properties of the bolometric comparison sample}
\label{tab:compbolsample}  
\centering
\smallskip
%\footnote
\begin{tabular}{l c c c c c c c c c}
\hline \hline
                 &   SNe    & M$_{Ni}$        & M$_{ej}$      & E$_{k}$           & Reference$^\dagger$  \\
                      &     & M$_{\odot}$     & M$_{\odot}$   & 10$^{51}$ erg     &     \\
\hline
                 IIb &  SN 1993J  & 0.10-0.14 & 1.3-3.5 & 0.7-1.4 & 1,2   \\
                     &   SN 2008ax  & 0.07-0.15 & 2-5  & 1-6 &  2,3,4 \\
                     &   SN 2011dh &  0.05-0.10 & 1.8-2.5 & 0.6-1.0 & 5  \\
                     &   SN 2011fu  & 0.15  & 3-5 & 1.3 &  6,16     \\
                     &   SN 2013df  & 0.10-0.13  & 0.8-1.4  & 0.4-1.2 & 7,16  \\
                     &   SN 2015as & 0.08 & 1.1-2.2 & 0.65 & 8  \\
                     &   SN 2020acat &  0.13$\pm$0.03 & 2.3$\pm$0.4  & 1.2$\pm$0.3       & 9   \\
\hline\hline
                Ib     &   SN 1999dn  & 0.11 & 4.0-6.0 & 5.0-7.5 & 10  \\
                     &   SN 2007Y   & 0.06$\pm$0.02 & 0.44 & 0.043 & 11 \\
                     &   SN 2009jf  & 0.17$\pm$0.03 & 4-9 & 3-8 & 12 \\
                     &   iPTF13bvn  & 0.06-0.09 & 2 & 1 & 13 \\     
                     &   SN 2015ap  & 0.01  & 3.90 & --  & 14 \\
                     &   SN 2017iro & 0.05-0.10 & 1.4-4.3 & 0.8-1.9 & 15 \\
\hline		                                                                                         
\end{tabular}
\newline
$^\dagger$ REFERENCES.-- (1)\cite{1995A&AS..110..513B}; (2) \cite{2016PhDT.......113M}; (3)\cite{2008MNRAS.389..955P}; (4)\cite{2011MNRAS.413.2140T}; (5)\cite{2013MNRAS.433....2S}; 
  (6)\cite{2013MNRAS.431..308K}; (7)\cite{2014MNRAS.445.1647M}; (8)\cite{2018MNRAS.476.3611G}; (9)\cite{2022MNRAS.513.5540M}; (10)\cite{2011MNRAS.411.2726B}; (11)\cite{2009ApJ...696..713S}; (12)\cite{2011MNRAS.416.3138V}; (13)\cite{2014MNRAS.445.1932S}; (14) \cite{2020MNRAS.497.3770G}; (15) \cite{2022ApJ...927...61K}; (16)\cite{2011ApJ...741...97D}
%\label{tab:photometric_parameters_different_SNe}.      
\end{table}

	\section{Radio Observations}
	\label{radio}
	
	SN~2022crv was first observed at radio wavelengths with the Australia Telescope Compact Array (ATCA) on 2022-03-01 UT \citep{ryder2022a,ryder2022b}, and radio emission was clearly detected at both 9.0 and 5.5 GHz. Radio monitoring with the ATCA at these 2 frequencies has continued for over a year, and the results are reported in Table~\ref{tab:radio}. Frequent observations of the nearby source PKS B0919-260 allowed us to monitor and correct for variations in gain and phase for each run. The data for each 2~GHz bandwidth have been edited and calibrated using tasks in the {\sc miriad} software package \citep{miriad}. Robust weighting was employed in imaging the visibilities, and after multifrequency synthesis and deconvolution, the flux densities were obtained from Gaussian fitting to the elliptical beam shape. As the ATCA primary flux calibrator PKS B1934-638 was sometimes not accessible during the observation periods, the relatively stable PKS B0823-500 was observed instead. This source is routinely monitored by ATCA staff, and by comparing its measured interpolated flux density with that measured from an image of the source on the day of observation, the flux densities for SN~2022crv have been placed on a uniform flux scale regardless of the flux calibrator adopted.
	
	We also carried out radio observations of SN~2022crv with the upgraded Giant Metrewave Radio Telescope (uGMRT) from 2022-03-31.58 UT to 2022-12-21.04 UT at multiple epochs. The observations were done in band-3 (250$-$500 MHz), band-4 (550$-$950 MHz), and band-5 (1050$-$1450 MHz). The data were recorded in the standard continuum mode with an integration time of 10 seconds. We used 200 MHz bandwidth in band-3 and 400 MHz bandwidth in bands-4 and 5, split into 2048 channels. 3C147 was used as the flux density calibrator, and J0837$-$198 was used as the phase calibrator. We used the Astronomical Image Processing system \citep[AIPS;][]{greisen2003} to analyze the uGMRT data and followed standard procedures from \citet{nayana2017}. The calibrated visibilities of the target source were imaged using AIPS task IMAGR. We performed a few rounds of phase-only self-calibration to improve image quality. The flux density was determined by fitting a two-dimensional Gaussian at the SN~position using AIPS task JMFIT. We present the details of uGMRT observations and flux densities in Table \ref{tab:radio}.

	\begin{figure}
		\centering   
		\resizebox{\hsize}{!}{\includegraphics{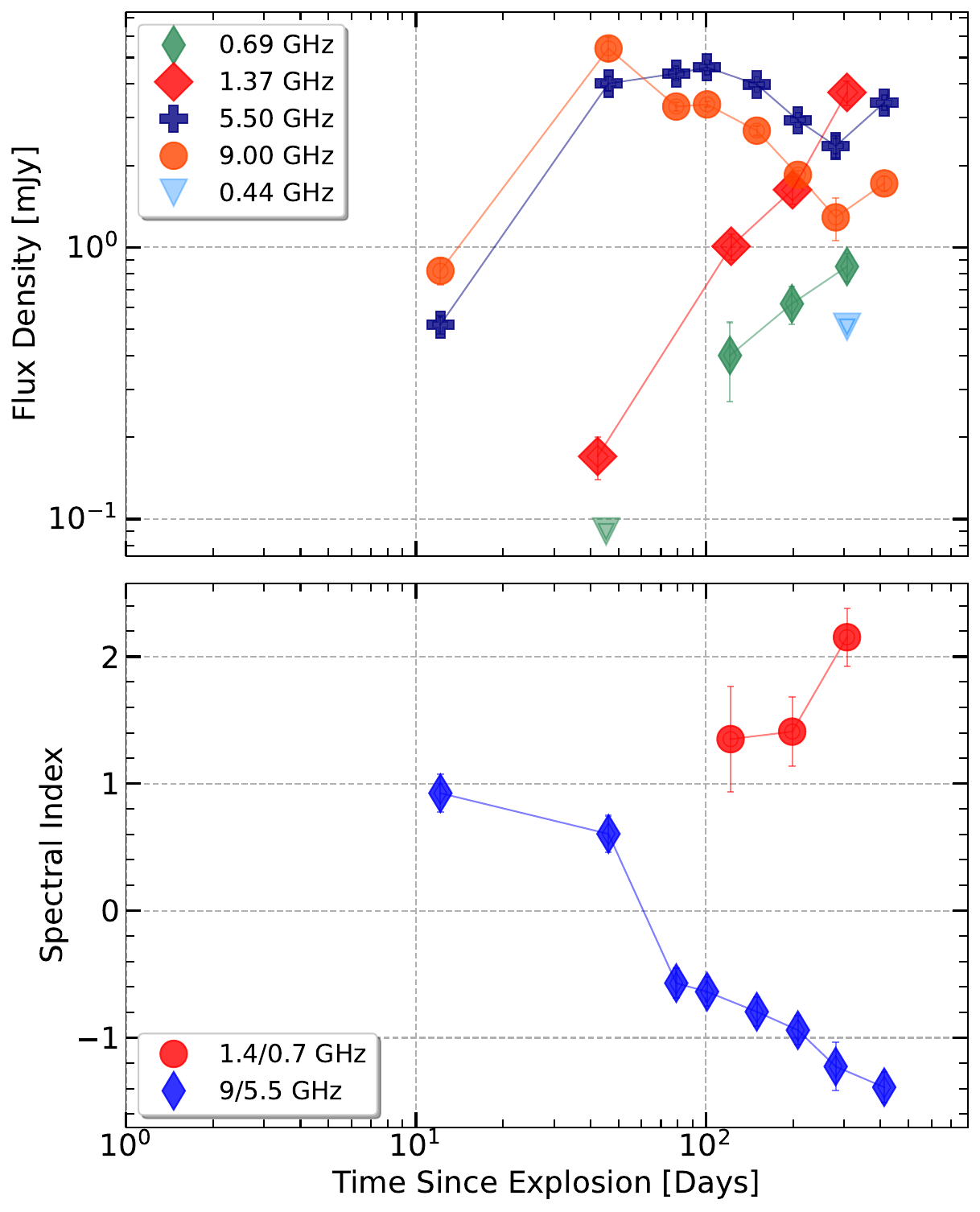}}
		\caption{Upper panel: Radio light curves of SN~2022crv at 0.44, 0.69, 1.37, 5.5, and 9.0 GHz. The down-pointing triangles are the upper limits of the flux densities. Lower panel: Near simultaneous spectral indices between 9/5.5 GHz (red circles) and 1.4/0.7 GHz (green circles).}
		\label{fig:radioplot}
	\end{figure}
	
	\begin{figure*}
		\centering   
		\resizebox{\hsize}{!}{\includegraphics{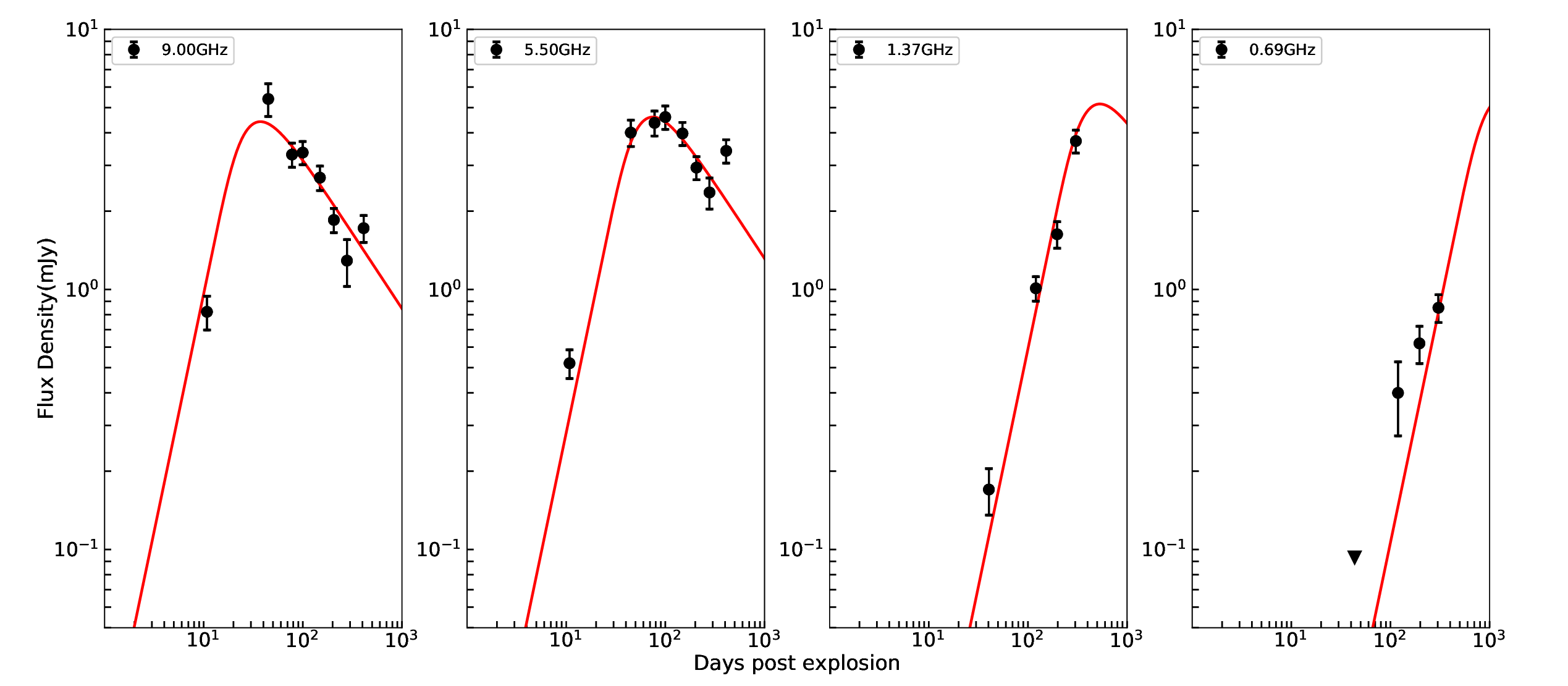}}
		\caption{Radio light curves of SN~2022crv at frequencies $\nu$\,=\,0.69 $-$ 9 GHz. The solid red curves represent the best-fit SSA model. The filled black circles denote the observed flux densities.}
		\label{fig:ssafit}
	\end{figure*}
	
	\subsection{Radio light curves and spectral indices}
	
	We detect radio emission from SN~2022crv at frequencies from 0.69 to 9.0 GHz during $t \sim$\,12 $-$ 412 d. The flux densities initially rise at all frequencies, reaching a peak spectral luminosity at 5.5 GHz of 6.51 $\times$ 10$^{27}$ erg\,s$^{-1}$\, Hz$^{-1}$ at $t \sim$ 100 d. The near-simultaneous spectral index, $\alpha$ ($F \propto \nu^\alpha$) between 9 GHz and 5.5 GHz is 0.92 $\pm$ 0.15 at $t \sim$ 12 d and approaches a value of $-$\,1.3 by $t \sim$ 412 d as the light curve transitions from the optically thick to the thin regime. The 1.4/0.7 GHz spectral indices are $\alpha$\,=\,1.35$\pm$0.42, 1.41$\pm$0.27, and 2.15 $\pm$ 0.23 at t $\sim$ 123, 200, and 308 d, respectively, flatter than the standard optically-thick limit (5/2). This can be attributed to the inhomogeneities in the magnetic field and/or relativistic electron distribution in the emitting region \citep{2017ApJ...841...12B,2021ApJ...912L...9N,chandra2019}. Radio light curves and the evolution of radio spectral indices are shown in Figure~\ref{fig:radioplot}.
	
	\subsection{Radio emission model}
	
	In the CSM interaction model of radio SNe, the emission is associated with a forward shock created as the SN~ejecta interacts with the wind from the progenitor star before the explosion \citep{1982ApJ...259L..85C}. At the shock, electrons are accelerated to relativistic velocities in amplified magnetic fields and emit synchrotron radiation. A fraction of the post-shock energy density is distributed into magnetic fields ($\epsilon_{\rm B}$) and relativistic electrons ($\epsilon_{\rm e}$), which are assumed to be constant throughout the evolution of the ejecta. The low-frequency emission is significantly suppressed by an absorption component. The absorption can be either free-free absorption (FFA) due to the ionized wind material along the line of sight \citep{1986ApJ...301..790W} or synchrotron self-absorption (SSA) due to the relativistic electrons that generate radio emission \citep{1998ApJ...499..810C}. The radio flux density initially rises rapidly and then declines, tracing the transition from the optically thick to the thin regime.
	
	We modeled the radio light curves with a standard SSA model \citep{1998ApJ...499..810C}, as the SSA likely dominates for a typical situation found for SE-SNe \citep{2006ApJ...651..381C}. The spectral and temporal evolution of radio flux densities $F(\nu,t)$ is given by 
	
	\begin{equation}
		\label{eqn:ssa-general-flux}
		F(\nu,t) = K_{1}\left(\frac{\nu}{5\, \rm GHz}\right)^{2.5} \left(\frac{t}{10\, \rm days}\right)^{a} \left[1-e^{-\tau_{\rm SSA}}\right]
	\end{equation}
	
	\begin{equation}
		\label{eqn:ssa-optical-depth}
		\tau_{\rm SSA} = K_{2}\left(\frac{\nu}{5\, \rm GHz}\right)^{-(p+4)/2}\left(\frac{t}{10\, \rm days}\right)^{-(a+b)}
	\end{equation}
	
	\noindent
	In the above equations, $K_{1}$ and $K_{2}$ are the flux density and optical depth normalization constants, respectively; $\tau_{\rm SSA}$ represents the optical depth due to synchrotron self-absorption; $a$ and $b$ denote the temporal indices of radio flux densities in the optically thick and thin regime, respectively; and $p$ is the power-law index of the relativistic electron energy distribution ($N(E) \propto E^{-p}$). We model the radio light curves keeping $K_{1}$, $K_{2}$, $a$, $b$, and $p$ as free parameters. We use the Markov Chain Monte Carlo (MCMC) method and choose 32 walkers and 5000 steps to estimate the best-fit values. We execute the fit using the Python package emcee \citep{2013PASP..125..306F}. The best-fit values of the parameters are $K_{1} =$ 0.22$^{+0.02}_{-0.02}$, $K_{2} =$ 93$^{+22}_{-18}$, $a =$ 1.80$^{+0.04}_{-0.04}$, $b =$ 0.58$^{+0.06}_{-0.06}$, and $p =$ 2.80$^{+0.27}_{-0.26}$. We present the best-fit model and the observed flux densities in Figure~\ref{fig:ssafit}. The corner plot showing how well the parameters are constrained is shown in Figure~\ref{fig:cornerplot}. 
	
	One can derive the shock radius ($R_{\rm s}$) and magnetic fields ($B$) using the peak frequency ($\nu_{\rm p}$) and peak flux density ($F_{\rm p}$) in the SSA scenario \citep[using equations 13 and 14 of][]{1998ApJ...499..810C}. We use $\nu_{\rm p} =$ 9 and 5.5 GHz and $F_{\rm p}$ from the best-fit modeled light curves to derive $R_{\rm s} =$ (1.07 $\pm$ 0.11) $\times$ 10$^{16}$ cm and (1.79 $\pm$ 0.19) $\times$ 10$^{16}$ cm at $\sim$ 37 and 75 d, respectively. The corresponding mean shock velocity ($R/t$) is $v \sim$ 0.1 c. The post-shock magnetic fields are $B =$ 0.87 $\pm$ 0.02 and 0.53 $\pm$ 0.01 G, at $\sim$ 37 and 75 d, respectively. We also estimate the mass-loss rate to be $\dot{M} \sim$ (1.9$-$2.8) $\times$ 10$^{-5}$ M$_{\odot}$ yr$^{-1}$ at these epochs from the magnetic field scaling relation \citep[equation 19 of][]{1998ApJ...499..810C} for a wind velocity of $v_{\rm w} \sim$ 1000 km\,s$^{-1}$ (typical of compact WR stars) and $\epsilon_{\rm B} =$ 0.33.

	\section{Progenitor of SN~2022crv}
	\label{progenitor}
	
	\begin{figure}
		\begin{center}
			\includegraphics[width=\columnwidth]{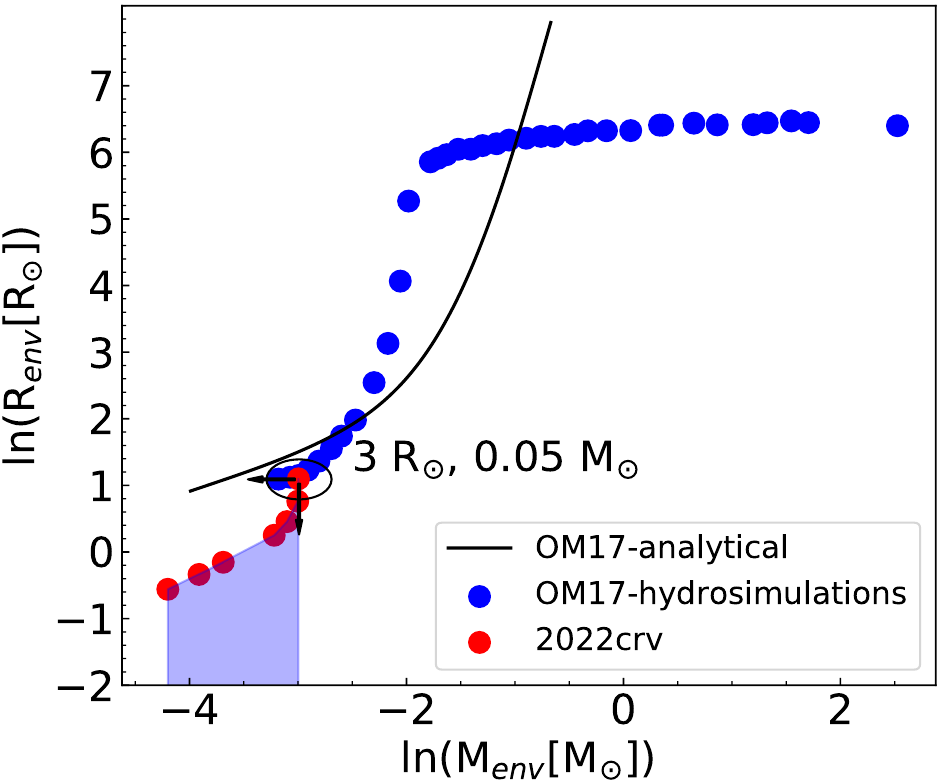}
		\end{center}
		\caption{The plot shows the relation between the radius and mass of the envelope for the SNe~IIb progenitor models (blue dots) generated by \cite{2017ApJ...840...90O}. The analytical relation derived by \citet{2017ApJ...840...90O} that explains the properties of the numerical evolution models is shown by the black line. Also shown here are the possible combinations of the radius and envelope mass of SN~2022crv (red dots) generated from the semi-analytical model of \citet{2016A&A...589A..53N}; note that the radius here is an upper limit for a given mass, and therefore the blue-shaded area is the allowed region for SN 2022crv. 
		}
		\label{fig:maedaouchi}
	\end{figure}
	
	In Section~\ref{bol}, we employed the \cite{2016A&A...589A..53N} bolometric light curve modeling to provide the upper limit for the envelope radius (R$_{\rm env}$) as a function of the envelope mass (M$_{\rm env}$). The best-fit mass is derived to be 0.015--0.05 M$_{\odot}$ and the radius is constrained to be $<$ 1--3 R$_{\odot}$. The constraint is shown in Figure \ref{fig:maedaouchi}. In this section we check whether the derived range of the envelope properties is consistent with the stellar evolution theory, and then provide a further constraint on the envelope properties assuming that the nature of the progenitor is explained by existing stellar evolution models. 
	
	\cite{2017ApJ...840...90O} (OM17) calculated a grid of binary evolution models for SNe~IIb. They further provide a sequence of single-star evolution models that can mimic the binary evolution scenario. In the following, we use this single-star (pseudo-binary-star) model sequence for discussion. Figure~\ref{fig:maedaouchi} shows the relationship between the radius and mass of the envelope for their SN IIb models (blue circles). The key property here is that the radius decreases as the envelope mass decreases below ln(M$_{\rm env}$[M$_{\odot}$] $\sim$ -2), as a result of an equilibrium configuration in the radiative envelope regime. OM17 also showed that this behavior can be approximately described analytically (solid black line), following an argument similar to that presented by \cite{1961ApJ...133..764C}. This analytic curve could be used to infer the model behavior in the very compact regime (R$_{\rm env}$ $\leq$ R$_{\odot}$) for which the numerical models are unavailable. 
	
	Our allowed range of R$_{\rm env}$ and M$_{\rm env}$ (see Figure~\ref{fig:maedaouchi}; blue shaded space) marginally overlaps with the model prediction only when the envelope properties are as follows; M$_{\rm env} \sim 0.05$ M$_\odot$ and R$_{\rm env} \sim 3$ R$_{\odot}$. Taking these as the most likely nature of the H-rich envelope attached to SN 2022crv, we conclude that the progenitor of SN 2022crv is very compact. The low estimated value of H envelope mass puts SN~2022crv in the category of the cSN~IIb class. The radius derived for SN 2022crv is among the smallest so far derived for SNe IIb, and overlaps with SNe Ib within the uncertainties of radii derived for individual objects. As such, SN 2022crv stands as the most compact progenitor for SNe IIb, representing a boundary between SNe IIb and Ib \citep{2018MNRAS.476.3611G,2020ApJ...903...70S,2022MNRAS.511..691G}.

	\begin{figure}
		\centering   
		\resizebox{\hsize}{!}{\includegraphics{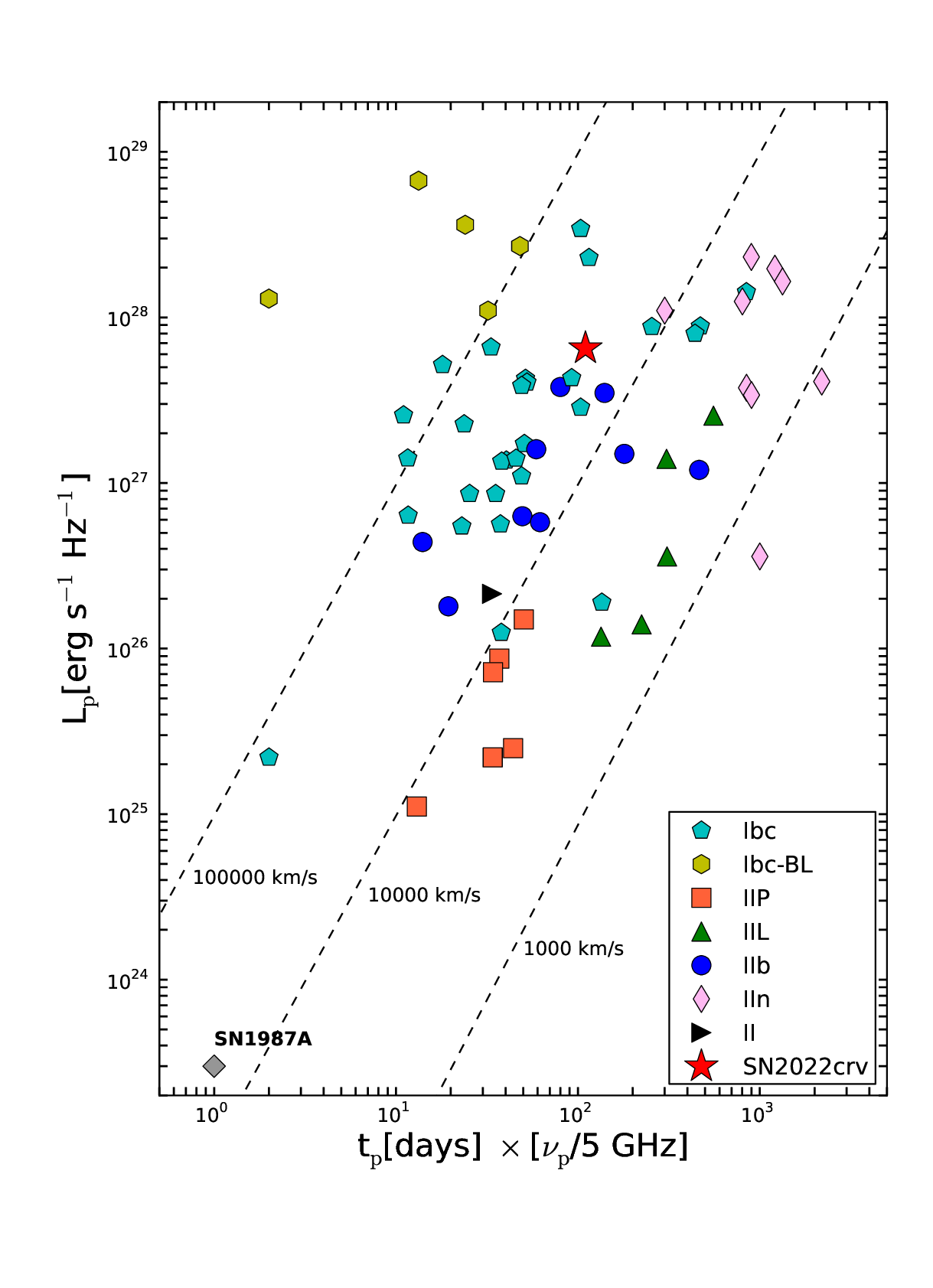}}
		\caption{The peak spectral luminosities versus time to peak for some well-observed CCSNe from the literature \citep[][and references therein]{bietenholz2021}. The position of SN\,2022crv is denoted as per the 5.5 GHz light curve. The dotted lines represent the mean velocities of radio-emitting shells in an SSA scenario with $p=3$.}
		\label{fig:snloc}
	\end{figure}
	
	We plot SN~2022crv (red star symbol) in the peak spectral luminosity $vs$ time to peak ($L_{\rm p}-t_{\rm p}$) diagram along with other CCSNe in Figure \ref{fig:snloc}. The dotted lines represent the mean shock velocities in the SSA scenario for $p=3$, assuming the equipartition of energy between relativistic electrons and magnetic fields ($\epsilon_{\rm e} = \epsilon_{\rm B} = 0.33$). \cite{chevalier2010} divided the radio-bright SNe IIb into two categories based on their radio properties: SNe cIIb (with compact progenitors), having faster shocks and less dense CSM; and SNe eIIb (with extended progenitors) having slower shocks and denser CSMs \citep{maeda2015}. This figure indicates that SN~2022crv is one of the radio-bright SN~IIb/Ib in the comparison sample. The position of SN 2022crv in the $L_{\rm p}-t_{\rm p}$ plane suggests that the SN~falls at a boundary between radio-bright SN~Ibc and SN~IIb; this agrees with the optical behavior of SN~2022crv transitioning from SNe~IIb to SNe~Ib, as explained in the previous sections.
	
\begin{figure*}
		\centering   
		\resizebox{\hsize}{!}{\includegraphics{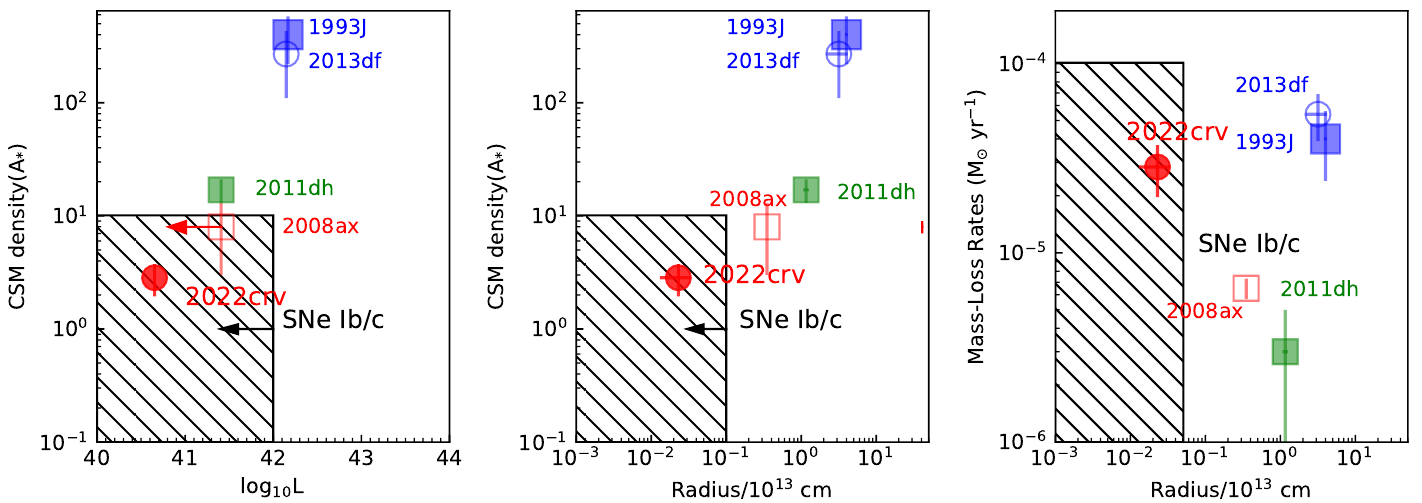}} 
		\caption{The relation between the pseudo-bolometric optical-NIR luminosity in the `shock-cooling phase' and the density of the CSM is shown in the first panel. Upper limits for the cooling luminosities are derived for the objects that lacked a primary peak, like SN~2008ax and SN~2022crv. The second panel shows the relation between the CSM density and the progenitor's radius, and the third panel describes the relation between mass-loss rates and the progenitor's radius. The typical CSM densities assumed for SNe~Ib/c are indicated by the shaded region \citep{2006ApJ...651..381C}, though they may contain a systematic uncertainty of up to an order of magnitude \citep{2012ApJ...758...81M}. 
		}
		\label{fig:rdm}
	\end{figure*}

	Figure~\ref{fig:rdm} displays how SN~2022crv is placed in the SN IIb/Ib/Ic progenitor property space. The CSM density is measured in terms of A$_{*}$ using the relation $\rho_{CSM}$ = 5 x 10$^{11}$ A$_{*}$ r$^{-2}$. The CSM density is estimated from the radio and X-ray combined analyses for SN 1993J \citep{1996ApJ...461..993F} and 2011dh \citep{2014ApJ...785...95M}; from the radio and optical combined analysis for SN 2013df \citep{maeda2015}; and from the radio data only for SN~2008ax \citep{chevalier2010}. The estimate of the CSM density for SNe Ib/c is based on radio data alone \citep{2006ApJ...651..381C}. For the comparison SNe IIb, the radii are taken from progenitor detection \citep{2017ApJ...840...90O}. On the other hand, the radii of SNe Ib/c are not strongly constrained (see \citealp{2012MmSAI..83..264M}). As the CSM densities and progenitor radii here do not represent a result of systematic analysis, we note that a substantial uncertainty is involved, and thus it should be taken as a demonstration. Further, additional uncertainty is introduced by the mass-loss wind velocity (assumed to be 20 km s$^{-1}$ for SNe~IIb and 1000 km s$^{-1}$ for SNe~Ib.), which has not been directly obtained for most of the samples. With these caveats in mind, \cite{maeda2015} formulated a relation between the early-phase cooling emission/progenitor radius and the CSM density/mass-loss rate. For SN 2022crv, we estimated our CSM density and mass-loss rate using the fits to the radio data of 5.5 GHz (t $\sim$ 75 d), assuming a wind velocity of 20 km s$^{-1}$. The progenitor radius for SN~2022crv has been constrained above (see Section~\ref{bol}).
	
	The first panel of Figure~\ref{fig:rdm} shows that the CSM density of SN 2022crv is on the higher end among the sample of SNe~Ib/c, similar to cSNe~IIb SNe~2008ax and 2011dh. For more extended progenitor objects like SNe~1993J and 2013df, the CSM densities are much higher than for cSNe~IIb, and about two orders of magnitude greater than for typical SNe~Ib/c. For the shock-cooling luminosity (the value plotted in the $x$-axis), the eSNe~IIb showed signatures of the shock-cooling phase, while for cSNe~IIb and Ib/c, only deep limits have been obtained \citep{2013ApJ...775L...7C}, including SN~2022crv for which we used the optical-NIR data to limit shock cooling luminosity. 
	
	The second panel of Figure~\ref{fig:rdm} relates the CSM densities with the progenitor's radius. The progenitor radius of $\sim$ 3 R$_{\odot}$ obtained for SN 2022crv fits into the relation among eSNe IIb, cSNe IIb, and SNe Ibc, i.e., higher CSM densities for more extended progenitors \citep{maeda2015}. The progenitor radius of SN~2022crv makes it one of the most compact SNe~IIb known so far, penetrating into the regime of SNe Ib/c.
	
	The third panel of Figure~\ref{fig:rdm} shows a more direct (but less certain) relationship between the nature of the progenitor and the mass-loss rates in the final stages of SN evolution. We observe a good overlap between the mass-loss rates of the SNe~IIb and SNe~Ib/c; indeed, the relation in the nature of the progenitors is diluted if one uses the mass-loss rate instead of the CSM density. A relatively high mass-loss rate inferred for SN 2022crv, despite its compact nature, further reverses the monotonic relation. Therefore, the relation between the progenitor radius and the mass-loss rate may be more complicated than postulated by \citet{maeda2015}. The binary interaction is believed to play a key role for these SNe~IIb (e.g., \citealt{2013ApJ...762...74B,2017ApJ...840...90O}) and thus the mass-loss rate is highly affected by the binary interaction (e.g., \citealt{2014ARA&A..52..487S}, for a review). Depending on the binary separation and mass ratio, the binary evolution could lead to a diversity in the mass-transfer history \citep[e.g.,][]{maeda2023b}.
	
	However, to further quantify the relations, especially the one related to the mass-loss rate, a more systematic approach will be required. The estimated CSM density can vary by orders of magnitudes when only the radio data are used (which is the case for SN 2022crv); the assumption of $\epsilon_{\rm e} = \epsilon_{\rm B} = 0.33$ adopted for the radio modeling of SN 2022crv is indeed very simplified and may be considered as an extreme assumption \citep{2021ApJ...918...34M,maeda2023}. Another issue is the wind velocity (v$_{\rm w}$), which is generally not well constrained (again, the case for SN 2022crv).

	\section{Summary and Conclusion}
	\label{sum}
	
	We present long-term photometric (up to +86 d) and spectroscopic (up to +33 d) observations of SN~IIb/Ib 2022crv. The spectral evolution of SN~2022crv implies that it is a SN~IIb that retained a thin H-envelope, showing a quick transition to SN Ib. The spectral evolution shows a prominent dip at 6200 \AA, which is well reproduced by \texttt{SYNAPPS} spectral modeling as a blend of Si {\sc ii} and H$\alpha$. The evolution of the EW and the wavelength of the absorption minimum of the feature (i.e., the velocity) also supports this identification as well as a quick transition from SN IIb to SN Ib. 
	
	The multi-band optical light curve shows the radioactive peak without strong shock-cooling emission, similar to what is seen for cSNe IIb. The absolute magnitude (M$_{V}$=$-$17.82$\pm$0.17 mag) and decay rate ($\Delta$m$_{15} (V)$=$0.76\pm0.04$) indicate SN~2022crv is a relatively bright and slowly-declining member of the SE-SNe sub-class. The bolometric light curve modeling inferred M$_{\rm Ni}$\,=\,0.12$\pm$0.05 M$_{\odot}$ and the ejecta mass in the range between 3.2--3.9 M$_{\odot}$. With the very early ATLAS data, we could place an upper limit on the radius of the envelope to be 3 R$_{\odot}$ and on the H envelope mass to be 0.05 M$_{\odot}$. Our observations of SN~2022crv show that it is one of the most compact SNe~IIb progenitors with a very thin H envelope (consistent with the recent reports by \citealp{2022MNRAS.511..691G}). Comparison with the SN IIb progenitor evolution models of \cite{2017ApJ...840...90O} shows that the upper limits, as mentioned above, most likely represent the envelope properties of SN 2022crv; it then represents one of the most compact SN IIb progenitors. This is a new constraint on the division of SN IIb and Ib progenitors.

	Radio observations of SN~2022crv spanning 0.44--9.0~GHz over a year indicated an interaction with a dense CSM. The SSA light curve modeling 
	provides the best-fit shock radius values to be  $R_{\rm s} =$ (1.07 $\pm$ 0.11) $\times$ 10$^{16}$ cm on day 37 and (1.79 $\pm$ 0.19) $\times$ 10$^{16}$ on day 75, indicating the shock velocity of $\sim 0.1$c. The estimated mass-loss rate is in the range of $\dot{M} \sim$ (1.9\,$-$\,2.8) $\times$ 10$^{-5}$ M$_{\odot}$ yr$^{-1}$, which however involves various uncertainties and thus should be regarded as a rough estimate. 
	In any case, the place of SN~2022crv in the radio luminosity phase diagram also predicts that the progenitor is compact, similar to SN~Ib/cSN~IIb. SN~2022crv is one of the radio-bright SN, having higher CSM densities than cSNe~IIb/Ib, indicating a high CSM density among the group of cSNe IIb and SNe Ib.

	The progenitor radius and the CSM density obtained for SN 2022crv fit into the relation among the progenitors of eSNe IIb, cSNe IIb, and SNe Ibc, i.e., higher CSM densities for more extended progenitors. On the other hand, in terms of the mass-loss rate, a high mass-loss rate inferred for SN 2022crv might be viewed as an outlier in the relation suggested so far (i.e., higher mass-loss rates for more extended progenitors). However, further quantifying this will require systematic analyses of the whole sample based on a uniform method, with a need to refine the treatment of the mass-loss wind velocity. Such investigation will be important to further understand the role of binary interaction toward SE-SNe. 
	
	While no CSM interaction signatures are seen in the optical spectra of SN 2022crv up to +33 d, we find interaction signatures in the radio waveband. 
	If a CSM is present but not extremely dense, then signatures in the optical become challenging to discern. There is a possibility that it might show up at later stages once the radioactive power decreases, as found for a few SNe IIb so far. Further observations in the years ahead are thus interesting, not only for SN 2022crv but for SE-SNe in general, to reveal whether any interaction signature emerges later.
	
	\section{Acknowledgments}
	The authors thank Melina Bersten for stimulating the discussion. K.M. acknowledges support from the Japan Society for the Promotion of Science (JSPS) KAKENHI grant JP18H05223, JP20H00174, and JP20H04737. This work was supported by Grant-in-Aid for Scientific Research (C), 22K03676. This work was supported by DST/JSPS, Grant number JPJSBP120227709. RAC acknowledges support from NSF grant AST-1814910. Nayana A.J. acknowledges DST-INSPIRE Faculty Fellowship (IFA20-PH-259) for supporting this research. RD acknowledges funds by ANID grant FONDECYT Postdoctorado Nº 3220449. B.A. acknowledges the Council of Scientific $\&$ Industrial Research (CSIR) fellowship award (09/948(0005)/2020-EMR-I) for this work. The data from the Seimei and Kanata telescopes were obtained under the KASTOR (Kanata And Seimei Transient Observation Regime) project, specifically under the following programs for the Seimei Telescope at the Okayama Observatory of Kyoto University (22A-K-0004, 22A-N-CT09). The Seimei telescope is jointly operated by Kyoto University and the Astronomical Observatory of Japan (NAOJ), with assistance provided by the Optical and Near-Infrared Astronomy Inter-University Cooperation Program. The authors thank the TriCCS developer team (supported by the JSPS KAKENHI grant Nos. JP18H05223, JP20H00174, and JP20H04736, and by NAOJ Joint Development Research). We thank the staff of IAO, Hanle, and CREST, Hosakote that made these observations possible. The facilities at IAO and CREST are operated by the Indian Institute of Astrophysics, Bangalore. We thank the staff of the GMRT that made these observations possible. GMRT is run by the National Centre for Radio Astrophysics of the Tata Institute of Fundamental Research. DKS and Nayana A.J. acknowledge the support provided by DST-JSPS under grant No. DST/INT/JSPS/P 363/2022.
	
	The Australia Telescope Compact Array is part of the Australia Telescope National Facility (https://ror.org/05qajvd42) funded by the Australian Government for operation as a National Facility managed by CSIRO. We acknowledge the Gomeroi people as the Traditional Owners of the Observatory site. This research made use of the NASA/IPAC Extragalactic Database (NED) that is operated by the Jet Propulsion Laboratory, California Institute of Technology, under contract with the National Aeronautics and Space Administration (NASA). Based in part on observations obtained at the 3.6\,m Devasthal Optical Telescope (DOT), which is a National Facility run and managed by Aryabhatta Research Institute of Observational Sciences (ARIES), an autonomous Institute under the Department of Science and Technology, Government of India. We thank the observers and operators at ST and DFOT facilities that made these observations possible.
	
	This research has made use of the APASS database, located on the AAVSO website. Funding for APASS has been provided by the Robert Martin Ayers Sciences Fund. This work has used data from the Asteroid Terrestrial-impact Last Alert System (ATLAS) project. The Asteroid Terrestrial-impact Last Alert System (ATLAS) project is primarily funded to search for near-earth asteroids through NASA grants NN12AR55G, 80NSSC18K0284, and 80NSSC18K1575; byproducts of the NEO search include images and catalogs from the survey area. This work was partially funded by Kepler/K2 grant J1944/80NSSC19K0112 and HST GO-15889, and STFC grants ST/T000198/1 and ST/S006109/1. The ATLAS science products have been made possible through the contributions of the University of Hawaii Institute for Astronomy, the Queen's University Belfast, the Space Telescope Science Institute, the South African Astronomical Observatory, and The Millennium Institute of Astrophysics (MAS), Chile. 
	
	\bibliography{refag}
	\bibliographystyle{aasjournal}
	
	\appendix
	
	\section{Log of Observations}
	
	Table \ref{tab:photstandard} lists the log of photometric standards that are used to calibrate the SN field.
	\begin{table*}
\centering
\caption{Log of Secondary Standard magnitudes for the field surrounding SN~2022crv. The magnitudes reported are in the Vega system.}
\label{tab:photstandard}
\begin{tabular}{c c c c c c}
\toprule
 ID &            $U$ &            $B$ &            $V$ &            $R$ &            $I$ \\
    &        (mag)   &      (mag)     &     (mag)      &      (mag)     &  (mag)         \\
\midrule
  1 & 17.49$\pm$0.16 & 17.53$\pm$0.17 & 17.04$\pm$0.10 & 16.64$\pm$0.15 & 16.26$\pm$0.17 \\
  2 & 16.83$\pm$0.15 & 16.42$\pm$0.16 & 15.74$\pm$0.10 & 15.30$\pm$0.14 & 14.91$\pm$0.16 \\
  3 & 17.06$\pm$0.15 & 16.57$\pm$0.16 & 15.87$\pm$0.10 & 15.38$\pm$0.14 & 14.97$\pm$0.16 \\
  4 & 17.25$\pm$0.16 & 17.09$\pm$0.16 & 16.55$\pm$0.10 & 16.15$\pm$0.15 & 15.77$\pm$0.17 \\
  5 & 15.02$\pm$0.15 & 15.01$\pm$0.16 & 14.48$\pm$0.10 & 14.12$\pm$0.14 & 13.79$\pm$0.16 \\
  6 & 15.65$\pm$0.15 & 14.38$\pm$0.16 & 13.41$\pm$0.10 & 12.75$\pm$0.14 & 12.19$\pm$0.16 \\
  7 & 16.99$\pm$0.15 & 16.79$\pm$0.16 & 16.10$\pm$0.10 & 15.69$\pm$0.14 & 15.31$\pm$0.16 \\
  8 & 17.23$\pm$0.15 & 16.97$\pm$0.16 & 16.36$\pm$0.10 & 15.94$\pm$0.14 & 15.55$\pm$0.16 \\
  9 & 16.33$\pm$0.15 & 16.10$\pm$0.16 & 15.50$\pm$0.10 & 15.11$\pm$0.14 & 14.75$\pm$0.16 \\
 10 & 16.80$\pm$0.15 & 16.57$\pm$0.16 & 16.00$\pm$0.10 & 15.57$\pm$0.14 & 15.20$\pm$0.16 \\
 11 & 17.41$\pm$0.16 & 17.33$\pm$0.16 & 16.63$\pm$0.10 & 16.21$\pm$0.14 & 15.78$\pm$0.16 \\
 12 & 16.21$\pm$0.15 & 15.95$\pm$0.16 & 15.31$\pm$0.10 & 14.92$\pm$0.14 & 14.58$\pm$0.16 \\
 13 & 17.01$\pm$0.15 & 17.00$\pm$0.16 & 16.48$\pm$0.10 & 16.16$\pm$0.14 & 15.83$\pm$0.16 \\
\bottomrule
\end{tabular}
\end{table*}
	The long-term temporal evolution of SN~2022crv has been carried out using a number of telescopes from India and Japan. The complete log of optical and NIR observations for the SN are logged in Table \ref{tab:photopt} and Table \ref{tab:photnir}.
	
	\begin{table*}
\centering
\caption{Log of Optical observations of SN~2022crv from 1.5m KT, 1m ST, 2m HCT-HFOSC, 1.3m DFOT and 3.8m Seimei Telescope. The magnitudes reported are in the Vega system.}
\label{tab:photopt}
\begin{tabular}{c c c c c c c c}
\toprule
    JD & Phase$^a$ & Telescope &            $B$ &   $g$ &         $V$ &            $R$ &            $I$ \\
(245 9600+) &   (d) &          &     (mag)      &  (mag) &     (mag)    &      (mag)     &     (mag)      \\
\midrule
36.19 &     -8.61 &    Seimei &            --- & 16.38$\pm$0.02 &      --- & 15.88$\pm$0.11 & 15.84$\pm$0.07 \\
 38.66 &     -6.14 &        KT &            --- & --- & 15.79$\pm$0.04 & 15.62$\pm$0.04 & 15.48$\pm$0.04 \\
 41.61 &     -3.19 &        KT &            --- & --- & 15.58$\pm$0.05 & 15.37$\pm$0.05 & 15.27$\pm$0.05 \\
 42.65 &     -2.15 &      DFOT & 16.18$\pm$0.10 & --- & 15.52$\pm$0.06 & 15.37$\pm$0.07 & 15.23$\pm$0.05 \\
 44.19 &     -0.61 &      DFOT & 16.16$\pm$0.09 ---& 15.50$\pm$0.05 & 15.30$\pm$0.07 & 15.13$\pm$0.10 \\
 45.09 &     +0.29 &        KT & 16.21$\pm$0.05 & --- & 15.51$\pm$0.04 & 15.28$\pm$0.04 & 15.12$\pm$0.05 \\
 45.12 &     +0.32 &     Seimei &           --- & 15.67$\pm$0.02 &          ---   & 15.23$\pm$0.10 & 15.28$\pm$0.11 \\
 46.62 &     +1.82 &        KT & 16.27$\pm$0.05 & --- & 15.52$\pm$0.05 & 15.22$\pm$0.04 & 15.08$\pm$0.05 \\
 47.59 &     +2.79 &       HCT & 16.36$\pm$0.05 & --- & 15.57$\pm$0.06 & 15.25$\pm$0.07 & 15.03$\pm$0.05 \\
 50.63 &     +5.83 &        KT &            --- & --- & 15.63$\pm$0.12 &            --- & 15.04$\pm$0.06 \\
 53.19 &     +8.39 &       HCT &            --- & --- & 15.77$\pm$0.06 & 15.38$\pm$0.08 & 15.14$\pm$0.10 \\
 54.59 &     +9.79 &        KT &            --- & --- & 15.95$\pm$0.10 & 15.46$\pm$0.08 &            --- \\
 62.62 &    +17.82 &        KT &            --- & --- & 16.48$\pm$0.05 & 15.86$\pm$0.05 & 15.48$\pm$0.06 \\
 67.19 &    +22.39 &      DFOT & 17.90$\pm$0.14 & --- & 16.77$\pm$0.08 & 16.08$\pm$0.07 & 15.62$\pm$0.12 \\
 68.09 &    +23.29 &    Seimei &            --- & 17.44$\pm$0.09 &            --- & 16.16$\pm$0.13 & 15.69$\pm$0.10 \\
 70.55 &    +25.75 &        KT &            --- & --- & 16.79$\pm$0.06 & 16.17$\pm$0.06 & 15.76$\pm$0.06 \\
 72.09 &    +27.29 &       HCT & 18.06$\pm$0.06 & --- & 16.86$\pm$0.04 & 16.24$\pm$0.07 & 15.82$\pm$0.10 \\
 73.39 &    +28.59 &      DFOT &            --- & --- & 16.94$\pm$0.07 & 16.35$\pm$0.07 & 15.88$\pm$0.09 \\
 73.58 &    +28.78 &        KT &            --- & --- & 16.96$\pm$0.05 & 16.31$\pm$0.07 & 15.87$\pm$0.05 \\
 75.62 &    +30.82 &        KT &            --- & --- & 16.97$\pm$0.06 & 16.36$\pm$0.06 & 15.85$\pm$0.08 \\
 88.50 &    +43.70 &        KT &            --- & --- & 17.26$\pm$0.04 & 16.68$\pm$0.05 & 16.21$\pm$0.04 \\
 90.09 &    +45.29 &      DFOT & 18.20$\pm$0.10 & --- & 17.33$\pm$0.06 &            --- &            --- \\
 90.49 &    +45.69 &      DFOT &            --- & --- &            --- & 16.76$\pm$0.07 & 16.26$\pm$0.08 \\
 91.49 &    +46.69 &        KT &            --- & --- & 17.26$\pm$0.07 & 16.74$\pm$0.06 & 16.26$\pm$0.05 \\
 94.09 &    +49.29 &        ST &            --- & --- & 17.35$\pm$0.09 & 16.80$\pm$0.08 &            --- \\
 96.69 &    +51.89 &      DFOT & 18.37$\pm$0.11 & --- & 17.46$\pm$0.06 & 16.94$\pm$0.10 & 16.39$\pm$0.10 \\
100.19 &    +55.39 &      DFOT &            --- & --- & 17.53$\pm$0.08 & 16.96$\pm$0.08 & 16.41$\pm$0.08 \\
101.99 &    +57.19 &    Seimei &            --- & 18.99$\pm$0.10 &            --- & 16.99$\pm$0.12 & 16.61$\pm$0.09 \\
102.47 &    +57.67 &        KT &            --- & --- & 17.46$\pm$0.06 & 16.94$\pm$0.04 & 16.46$\pm$0.05 \\
119.39 &    +74.59 &      DFOT &            --- & --- &            --- & 17.30$\pm$0.09 & 16.72$\pm$0.11 \\
120.39 &    +75.59 &      DFOT &            --- & --- &            --- & 17.23$\pm$0.07 & 16.75$\pm$0.10 \\
124.48 &    +79.68 &        KT &            --- & --- & 18.00$\pm$0.06 & 17.31$\pm$0.11 & 16.71$\pm$0.07 \\
130.47 &    +85.67 &        KT & 18.86$\pm$0.06 & --- & 17.99$\pm$0.19 & 17.33$\pm$0.09 & 16.91$\pm$0.09 \\
\bottomrule
\multicolumn{3}{l}{$^a$\footnotesize{Time since $V$-Band Maximum}}
\end{tabular}
\end{table*}
	\begin{table}
\centering
\caption{Log of Near-Infrared observations of SN~2022crv from HONIR mounted on 1.5m KT. The magnitudes reported are in Vega system.}
\label{tab:photnir}
\begin{tabular}{c c c c c}
\toprule
    JD      & Phase$^*$ &            $J$ &            $H$ &           $Ks$ \\
(245 9600+) &   (d)     &        (mag)   &      (mag)     &     (mag)      \\

\midrule
 38.66 &     -6.14 & 15.22$\pm$0.02 & 15.08$\pm$0.04 & 14.96$\pm$0.04 \\
 41.61 &     -3.19 & 15.02$\pm$0.02 & 14.84$\pm$0.03 & 14.63$\pm$0.04 \\
 42.65 &     -2.15 & 14.96$\pm$0.02 & 14.77$\pm$0.03 & 14.53$\pm$0.04 \\
 44.62 &     -0.18 & 14.87$\pm$0.02 & 14.70$\pm$0.03 & 14.41$\pm$0.04 \\
 46.62 &     +1.82 & 14.79$\pm$0.02 & 14.62$\pm$0.02 & 14.31$\pm$0.03 \\
 50.65 &     +5.58 &            --- &            --- & 14.28$\pm$0.10 \\
 54.59 &    +9.79 & 14.79$\pm$0.04 & 14.69$\pm$0.10 & 14.25$\pm$0.07 \\
 62.62 &    +17.82 & 15.05$\pm$0.02 & 14.67$\pm$0.03 & 14.46$\pm$0.05 \\
 70.55 &    +25.75 & 15.36$\pm$0.03 & 15.12$\pm$0.07 & 14.74$\pm$0.07 \\
 73.58 &    +28.78 & 15.49$\pm$0.03 & 14.99$\pm$0.03 & 14.80$\pm$0.04 \\
 75.62 &    +30.82 & 15.59$\pm$0.03 & 14.99$\pm$0.12 & 14.84$\pm$0.10 \\
 88.50 &    +43.70 & 16.04$\pm$0.03 & 15.40$\pm$0.04 & 15.31$\pm$0.21 \\
 91.49 &    +46.69 & 16.32$\pm$0.07 & 15.54$\pm$0.19 & 15.31$\pm$0.22 \\
102.47 &    +57.67 & 16.39$\pm$0.04 & 15.75$\pm$0.04 & 15.62$\pm$0.07 \\
\bottomrule
\multicolumn{3}{l}{$^a$\footnotesize{Time since V-Band Maximum}}
\end{tabular}
\end{table}
	
	The log of spectroscopic observations of SN~2022crv showing a coverage up to +33 d post maximum is shown in Table \ref{tab:2022crv_spec_obs}.
	\begin{table}
\caption{Log of spectroscopic observations of SN~2022crv. The phase is measured with respect to $V$-band maximum).}
\label{tab:2022crv_spec_obs}
\begin{center}
%\smallskip
%\small\addtolength{\tabcolsep}{-2pt}
\begin{tabular}{c c c c c c}
\hline \hline
Phase  &    Telescope  &     Instrument  &		Range \\
       &               &                 &             \AA   \\
\hline    
-15.3      &  Gemini-North    & GMOS                    &   3500-7000           \\
-9.7      &   Seimei    &    KOOLS-IFU                  &   4100-8900   \\
0.4      &    HCT       &    HFOSC                      &   3800-6840             \\
0.2      &    Seimei    &    KOOLS-IFU                  &  4100-8900   \\
3.5     &     HCT         &   HFOSC                     &  3800-6840\\
6.2      &    Seimei   &     KOOLS-IFU                  &   4100-8900  \\
6.4     &     DOT    &       ADFOSC                     &   3500-8900   \\
8.3     &     HCT    &       HFOSC                      &   3800-8300 \\
23.2     &    Seimei    &    KOOLS-IFU                  &   4100-8900\\
33.2     &    Seimei    &    KOOLS-IFU                  &   4100-8900 \\
33.3     &    HCT   &        HFOSC                      &   3400-9500 \\
\hline
\end{tabular}
\end{center}
\end{table}

	Finally, the multi-frequency radio coverage of SN 2022crv covering from 0.69 GHz - 9 GHz is tabulated in Table \ref{tab:radio}.
	
	\begin{deluxetable}{lccccc}
		\tablecaption{Radio observations of SN~2022crv \label{tab:radio}}
		\tablehead{
			\colhead{Date of Observation} & \colhead{JD} & \colhead{Age\tablenotemark{a}} & \colhead{Frequency (GHz)} & \colhead{Flux density\tablenotemark{b} (mJy)}
		}
		\startdata
		\multicolumn{5}{c}{uGMRT} \\
		\hline
		2022 Mar 31.58 & 2459670.08  & +42.33 & 1.37 & 0.17 $\pm$ 0.03 \\
		2022 Apr 03.54 & 2459673.04 & +45.29 & 0.69 & $<$ 0.09 \\
		2022 Jun 18.35 & 2459748.85 & +121.10 & 0.69 & 0.40 $\pm$ 0.13 \\
		2022 Jun 19.47 & 2459749.97 & +122.22 & 1.37 & 1.01 $\pm$ 0.11 \\
		2022 Sep 03.30 & 2459825.80 & +198.05 & 0.69 & 0.62 $\pm$ 0.10 \\
		2022 Sep 04.26 & 2459826.76 & +199.01 & 1.37 & 1.63 $\pm$ 0.19 \\
		2022 Dec 20.87 & 2459934.37 & +306.62 & 1.37 & 3.72 $\pm$ 0.38 \\
		2022 Dec 20.96 & 2459934.46 & +306.71 & 0.69 & 0.85 $\pm$ 0.10 \\
		2022 Dec 21.04 & 2459934.54 & +306.79 & 0.44 & $<$ 0.51 \\
		\hline
		\multicolumn{5}{c}{ATCA} \\
		\hline
		2022 Mar 01.4 &  2459639.90  & +12.15 & 5.5  & 0.52 $\pm$ 0.04 \\
		2022 Apr 04.4 &  2459673.90  & +46.15 & 5.5  & 4.01 $\pm$ 0.24 \\
		2022 May 07.3 &  2459706.80  & +79.05 & 5.5  & 4.37 $\pm$ 0.20 \\
		2022 May 29.2 &  2459728.70  & +100.95 & 5.5  & 4.60 $\pm$ 0.12 \\
		2022 Jul 17.1 &  2459777.60  & +149.85 & 5.5  & 3.98 $\pm$ 0.08 \\
		2022 Sep 12.9 &  2459835.40  & +207.65 & 5.5  & 2.94 $\pm$ 0.06 \\
		2022 Nov 24.6 &  2459908.10  & +280.35 & 5.5  & 2.36 $\pm$ 0.22 \\
		2023 Apr 05.4 &  2460039.90  & +412.15 & 5.5  & 3.41 $\pm$ 0.09 \\
		2022 Mar 01.4 &  2459639.90  & +12.15 & 9.0  & 0.82 $\pm$ 0.09 \\
		2022 Apr 04.4 &  2459673.90  & +46.15 & 9.0  & 5.40 $\pm$ 0.56 \\
		2022 May 07.3 &  2459706.80  & +79.05 & 9.0  & 3.30 $\pm$ 0.12 \\
		2022 May 29.2 &  2459728.70  & +100.95 & 9.0  & 3.36 $\pm$ 0.08 \\
		2022 Jul 17.1 &  2459777.60  & +149.85 & 9.0  & 2.69 $\pm$ 0.11 \\
		2022 Sep 12.9 &  2459835.40  & +207.65 & 9.0  & 1.85 $\pm$ 0.07 \\
		2022 Nov 24.6 &  2459908.10  & +280.35 & 9.0  & 1.29 $\pm$ 0.23 \\
		2023 Apr 05.4 &  2460039.90  & +412.15 & 9.0  & 1.72 $\pm$ 0.11 \\
		\enddata
		\tablenotetext{a}{The age is calculated assuming 2459627.75 as the date of explosion.}
		\tablenotetext{b}{The uGMRT errors on the flux densities are the sum of errors from Gaussian fitting on the SN emission (using AIPS task JMFIT) and a 10\% calibration uncertainty added in quadrature.}
	\end{deluxetable}
	
\begin{figure}
		\centering   
		\resizebox{\hsize}{!}{\includegraphics{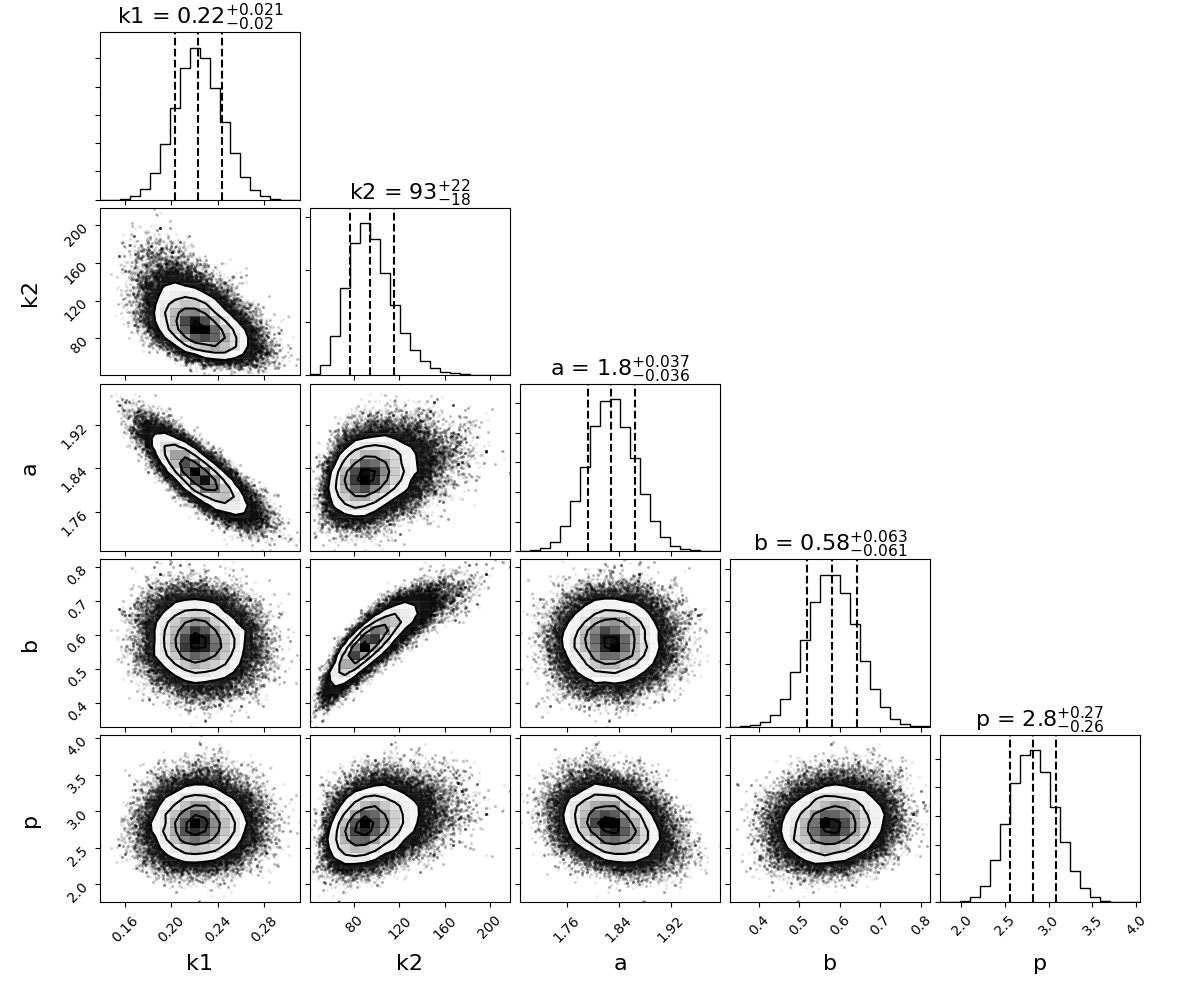}}
		\caption{The corner plot shows the results of MCMC modeling of the SN~2022crv radio data with the SSA model. The parameters here are according to Eq 3 and 4, respectively.}
		\label{fig:cornerplot}
	\end{figure}

	The corner plot showing the confidence of the parameters obtained from the SSA modeling of the radio data of SN~2022crv is shown in Figure \ref{fig:cornerplot}
	
\end{document}